\documentclass[preprint,tighten,twocolumn,onecolappendix]{aastex7}
\usepackage{apjfonts}
\usepackage{amsmath}
\usepackage{multirow}
\usepackage{textcomp}
\usepackage{appendix}
\usepackage{color}
\usepackage{graphicx}
\usepackage{hyperref}
\usepackage{subfig}

\shorttitle{ALMA Molecular Line Search of the M87 Nucleus}
\shortauthors{Boizelle et al.}

\usepackage{natbib}

\newcommand{\kms}{km s\ensuremath{^{-1}}}
\newcommand{\vlos}{\ensuremath{v_\mathrm{LOS}}}

\newcommand{\sigmalos}{\ensuremath{\sigma_\mathrm{LOS}}}

\newcommand{\vsys}{\ensuremath{v_\mathrm{sys}}}

\newcommand{\mbh}{\ensuremath{M_\mathrm{BH}}}
\newcommand{\msun}{\ensuremath{M_\odot}}

\newcommand{\per}{\ensuremath{^{-1}}}
\newcommand{\pertwo}{\ensuremath{^{-2}}}
\newcommand{\perthree}{\ensuremath{^{-3}}}

\newcommand{\coone}{CO(1\ensuremath{-}0)}
\newcommand{\cotwo}{CO(2\ensuremath{-}1)}
\newcommand{\cothree}{CO(3\ensuremath{-}2)}
\newcommand{\cofour}{CO(4\ensuremath{-}3)}

\begin{document}

\title{CO Emission and Absorption-line Survey of the M87 Nucleus Using Archival ALMA Imaging}

\author[0000-0001-6301-570X]{Benjamin D. Boizelle}
\affiliation{Department of Physics and Astronomy, N284 ESC, Brigham Young University, Provo, UT, 84602, USA}
\email{boizellb@byu.edu}

\author[0009-0008-1795-3576]{Xueyi Li}
\affiliation{Department of Physics \& Astronomy, Macalester College, 1600 Grand Avenue, Saint Paul, MN 55105, USA}
\affiliation{Department of Physics and Astronomy, University of Delaware, 104 The Green
Newark, DE, 19716, USA}
\email{xlilev@udel.edu}

\author[0009-0004-7059-8658]{Nicholas LeVar}
\affiliation{Department of Physics and Astronomy, N284 ESC, Brigham Young University, Provo, UT, 84602, USA}
\email{nal43@student.byu.edu}

\author[0009-0001-2512-1429]{Sam Norcross}
\affiliation{Department of Physics and Astronomy, N284 ESC, Brigham Young University, Provo, UT, 84602, USA}
\email{san307@student.byu.edu}

\author[0000-0003-4559-8088]{Benjamin J. Derieg}
\affiliation{Department of Physics and Astronomy, N284 ESC, Brigham Young University, Provo, UT, 84602, USA}
\affiliation{Department of Physics and Astronomy, 201 James Fletcher Bldg. University of Utah, Salt Lake City, UT, 84112, USA}
\email{benjamin.derieg@utah.edu}

\author[0000-0003-3900-6189]{Jared R. Davidson}
\affiliation{Department of Physics and Astronomy, N284 ESC, Brigham Young University, Provo, UT, 84602, USA}
\email{jaredrd2@student.byu.edu}

\author[0009-0004-6545-8511]{Kavin Siaw}
\affiliation{Department of Physics, 10 East 2nd South, ROM 118, Brigham Young University -- Idaho, Rexburg, ID, 83460, USA}
\email{sia21002@byui.edu}

\author[0000-0002-1881-5908]{Jonelle L. Walsh}
\affiliation{George P. and Cynthia Woods Mitchell Institute for Fundamental Physics and Astronomy, 4242 TAMU, Texas A\&M University, College Station, TX, 77843-4242, USA}
\email{walsh@tamu.edu}

\begin{abstract}
We present an M87 molecular line search from archival Atacama Large Millimeter/sub-millimeter Array (ALMA) imaging, covering the circumnuclear disk (CND) as well as ionized gas filaments and dusty cloud regions. We find no evidence for CO emission in the central $\sim$kpc and place an upper limit of $M_\mathrm{H_2} < 2.3\times 10^5$ \msun\ in the atomic gas CND region, a factor of 20$\times$ lower than previous surveys. During this search, we discovered extragalactic CO absorption lines in the $J$ = 1$-$0, 2$-$1, and 3$-$2 transitions against the bright (Jy-scale) active nucleus. These CO lines are narrow ($\sim$5 \kms) and blueshifted with respect to the galaxy's systemic velocity by $-$75 to $-$84 \kms. This CO absorber appears to be kinematically distinct from outflowing atomic gas seen in absorption. Low integrated opacities ranging from $\tau_\mathrm{CO} \sim 0.02-0.06$ \kms\ and a column density $N_\mathrm{CO} \approx (1.2\pm0.2)\times 10^{15}$ cm\pertwo\ translate to $N_\mathrm{H_2} \sim (1-2) \times 10^{20}$ cm\pertwo. CO excitation temperatures spanning $T_\mathrm{ex} \sim 8$ K to $\sim$30 K do not follow local thermodynamic equilibrium (LTE) expectations, and non-LTE \texttt{radex} radiative transfer modeling of the CO absorber is consistent with a number density $n_\mathrm{H_2} \sim 5000$ cm\perthree\ embedded in a $\sim$60 K environment. Taken together, the observed CO absorption lines are most consistent with a thin, pressure-confined filament seen slightly off-center from the M87 nucleus. We also explore the impact of residual telluric lines and atmospheric variability on narrow extragalactic line identification and demonstrate how bandpass calibration limitations may introduce broad but very low S/N and spurious absorption and emission signatures.
\end{abstract}


\section{Introduction}

Ongoing gas accretion onto supermassive black holes (BHs; with masses $\mbh \sim 10^5 - 10^{10}$ \msun) powers active galactic nuclei (AGN) at the centers of galaxies. In extreme cases, highly efficient central engines are visible across the observable universe. A wide range of AGN phenomena -- including broad and/or narrow emission lines, a very hot accretion disk, and large-scale jets of relativistic plasma -- are explained in the AGN standard model primarily by different viewing angles and accretion rates \citep[e.g.,][]{tadhunter08,trump11,bianchi12,heckman14}. A subset of these AGN are radio bright with collimated jet features, typically classified as either type I (centrally dominant or edge-darkened continuum emission) or type II (edge-brightened) Fanaroff-Riley galaxies \citep[FR;][]{fanaroff74}.

M87 (NGC 4486) is one of the most well-studied luminous early-type galaxies \citep[ETGs; classified as cD0-1 pec;][]{devauc91}, in large part due to its close proximity in the Virgo cluster and as the brightest galaxy of the Virgo A subgroup. This FR I galaxy is also well studied due to its central narrow-lined, low-ionized nuclear emission-line region \citep[LINER Type 2;][]{ho97,dopita97}. The active nucleus is bright across mm to radio wavelengths (with central flux density $S_\nu \gtrsim 1$ Jy) and launches jets seen primarily on the approaching side across radio to X-ray wavelengths \citep[e.g.,][]{marshall02,kim18}, connecting to radio lobes that extend out to a projected $\sim$3 kpc \citep{chiaberge99}. The mass accretion rate of M87 is very low, between $\dot{M}_\mathrm{BH}<10^{-3} - 10^{-2}$ \msun\ yr\per\ or an Eddington fraction $\dot{M}_\mathrm{BH}/\dot{M}_\mathrm{Edd} \lesssim (0.15-1.5)\times 10^{-3}$ \citep[][]{kuo14,inayoshi20}. Together with many other radio galaxies, M87 lies in a transition region between cold, thin disk accretion and radiatively inefficient accretion flows.

M87 garnered significant additional attention during the Event Horizon Telescope (EHT) very large baseline interferometry (VLBI) campaigns that imaged its BH shadow on $\sim$25 $\mu$as scales and also connected gas accretion and magnetic field structures very close to the BH and the jet launching region \citep{event19,event21a,event21b}. Modeling of both the EHT visibilities and separate stellar kinematics returned a consistent BH mass ranging between $(5.4-6.5)\times 10^9$ \msun\ \citep{gebhardt09,event19,liepold23,simon24}. Earlier gas-dynamical modeling efforts had preferred an \mbh\ only half as massive \citep{harms94,macchetto97,walsh13}, although recent work has brought the gaseous results into good agreement \citep{osorno23}.

Early ground-based and Hubble Space Telescope (HST) imaging revealed filaments of ionized gas extending out several kpc from the M87 nucleus \citep{sparks93,ford94,pogge00}. It is unclear if this gas is related to known merger activity \citep{longobardi15}, or even if the gas in the central few kpc is infalling towards the BH \citep[e.g., in chaotic cold-mode accretion, or CCA;][]{gaspari13} or is being disturbed by rising bubbles or the radio jet \citep[e.g.,][]{Olivares2019}. In some massive elliptical or cD galaxies like M87, spectral energy distribution (SED) fitting \citep{yuan14} suggests hot-mode accretion dominates most of the time, even if a gaseous circumnuclear disk (CND) is clearly detected.

Many FR I galaxies host morphologically regular, geometrically thin disks with relatively small radii \citep[generally a few $\times$100 pc;][]{dekoff00,saripalli12}. These CNDs are often identified by dusty disks or lanes, and molecular line surveys have revealed modest cospatial H$_2$ mass reservoirs \citep[with typical $M_\mathrm{H_2} \sim 10^7 - 10^8$ \msun; e.g.,][]{ocana10,prandoni10,boizelle17,boizelle21,ruffa19,tadhunter24}. Unfortunately, the bright optical AGN and jet of M87 prevent any definitive conclusions about dust absorption on the smallest scales \citep[with the current consensus being an obscuration-free nucleus;][]{pogge00}. On larger scales, its filamentary gas appears to connect to an ionized-gas CND with well-ordered rotation out to $R\sim0\farcs5$ \citep[<50 pc;][]{sparks93,ford94,walsh13,boselli19,osorno23}. Atomic gas line-of-sight (LOS) speeds (\vlos) for this rotation reach roughly $\pm 500$ \kms\ in the inner $R\lesssim 0\farcs 1$ from the galaxy's systemic velocity $\vsys \equiv cz_\mathrm{obs} = 1284\pm 5$ \kms\ from stellar absorption-line fits with a heliocentric $z_\mathrm{obs} = 0.004283$ \citep{cappellari11}.


Despite the mostly ordered central velocity field, maps of the atomic emission-line dispersion $\sigmalos$ show asymmetries in the northwest direction that suggest a bi-conical outflow with intrinsic speeds reaching out to $\sim$400 \kms\ along a position angle PA~$\sim -45\degr$ \citep{sparks93,boselli19,osorno23}. Further support for an outflow comes from UV/optical absorption lines detected against the bright nucleus, although this may reveal a kinematically distinct component. These atomic tracers show blueshifted absorption lines with $\vlos -\vsys \sim -131\pm 19$ \kms\ and moderate intrinsic FWHM~$\lesssim 300$ \kms\ \citep[][]{tsvetanov99b,sabra03}. Absorption-line measurements as well as modeling of the central X-ray spectrum both support hydrogen column densities in the $N_\mathrm{H} \sim 10^{19} - 10^{20.7}$ cm\pertwo\ range \citep{wilson02,dimatteo03,sabra03}.

\begin{figure*}[!ht]
    \centering
    \includegraphics[width=\textwidth]{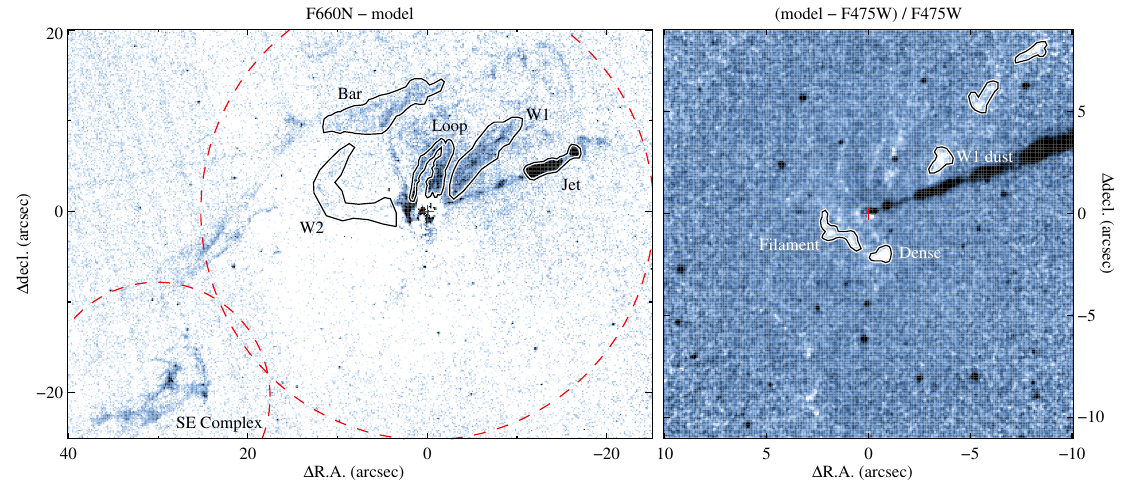}
    \caption{HST imaging of M87 after subtracting a stellar continuum MGE model. Narrowband residuals (\textit{left panel}) reveal H$\alpha$+[\ion{N}{2}] emission-line filaments above the low pattern noise that are found predominantly to the north of the nucleus (labeled with a +) and stretch out a few kpc. The most extended ALMA 12-m field of view (FoV) for imaging centered on the nucleus (half-power beam width HPBW~$\sim 50\arcsec$ in Band 3) and the FoV covering the southeast complex \citep[HPBW~$\sim 25\arcsec$ in Band 6;][]{simionescu18} are overlain (dashed circles). Broadband fractional differences (\textit{right}) reveal dusty clumps and filaments in the W1 and loop regions as well as in the SE direction. In the emission-line filament region W1, dust is identified in three main clumps. The dense and filament dust regions reach fractional deficits of 25\% and 10$-$15\%, respectively. Outlined regions shown here isolate the most obvious atomic gas emission or dust obscuration and are used in CO emission-line searches.}
    \label{fig:hst}
\end{figure*}

Within the central few kpc of M87 (out to at least $\sim$40\arcsec), molecular gas is best traced in the optical by discrete, low-extinction filaments and clouds. Similar dusty features are observed in the inner $\sim$kpc of most luminous elliptical galaxies in high-density environments \citep[][]{mathews03,temi18,tremblay18,Olivares2019}, often following ionized atomic emission-line filaments \citep[e.g.,][]{sparks93,ferrarese06,madrid07,boselli19,taylor20}. For M87, no central extinction is detected down to $R\sim 0\farcs15$ \citep[$\sim$10 pc;][]{prieto21}. Evidence for the corresponding thermal dust continuum is ambiguous and global SED fits are often interpreted as showing no evidence for diffuse dust and limited support for circumnuclear dust \citep{perlman01,xilouris04,buson09,baes10}. However, a couple of elevated SED points or mid-IR spectra \citep[][]{shi07,perlman07} have been tied to a cold dust component with characteristic temperature $T_\mathrm{dust} \sim 25-55$ K with mass $M_\mathrm{dust} < 7\times 10^4$ \msun\ \citep{baes10}.

Molecular gas that is expected to accompany the filamentary dust has remained elusive. Single-dish mm/sub-mm imaging of this target did not detect global $^{12}$CO (hereafter CO) emission in the $J=1-0$ transition across a velocity range of a few hundred \kms, first with an initial 1$\sigma$ line flux upper limit of $S_\mathrm{CO(1-0)}\Delta v < 20$ Jy \kms\ \citep[][]{bieging81,jaffe87,braine93} and later a 3$\sigma$ upper limit of 10.2 Jy \kms\ \citep[][]{combes07,salome08}. From interferometric observations with the Submillimeter Array (SMA), \citet{tan08} reported a tentative $S_\mathrm{CO(2-1)}\Delta v \approx 8.8\pm2.2(\mathrm{stat})\pm1.8(\mathrm{sys})$ Jy \kms\ detection (with separate 1$\sigma$ statistical and systematic uncertainties) in an $\sim$80 pc-diameter aperture about the nucleus. If correct, the corresponding total H$_2$ mass $M_\mathrm{H_2} \lesssim 5\times 10^6$ \msun\ could be detected using the Atacama Large Millimeter/submillimeter Array (ALMA), which provides an order-of-magnitude improvement in both angular resolution and limiting sensitivity over previous mm-wavelength interferometers. 


In this paper, we present a detailed mm/sub-mm wavelength molecular line search of the M87 nucleus using ALMA archival data. In a more brief study, \citet{ray24} analyzed the earliest two of these ALMA data sets and purport to detect and resolve central CO emission with $S_\mathrm{CO(1-0)}\Delta v = 3.85\pm0.4$ Jy \kms. The corresponding $M_\mathrm{H_2} \sim 10^7$ \msun\ is claimed to originate in a CND that extends out to at least $R\sim 100$ pc. They also identify absorption-line features against the bright nucleus. We cannot recover similar emission-line results, and nor do we support their absorption-line measurements. In our data calibration description and spectral measurements, we will reference their claims as needed.

This paper is structured as follows. In Section~\ref{sec:data_overview}, we introduce archival HST and ALMA data sets and imaging processes. Section~\ref{sec:line} describes the CO emission-line search and CO absorption-line detection. We constrain molecular gas properties in Section~\ref{sec:disc} and conclude in Section~\ref{sec:conc}. In the extensive appendices, we provide more in-depth analysis concerning extragalactic spectral line identification with ALMA, including the impact of atmospheric (ozone) variability and the possibility of spurious broad line detection due to bandpass calibration limitations.



Throughout, we assume a standard $\Lambda$CDM cosmology with $H_0 = 67.8$ \kms\ Mpc\per\ along with $\Omega_\mathrm{m} = 0.308$ and $\Omega_\Lambda = 0.692$ \citep{planck2016}. We adopt a distance modulus $m-M = 31.11\pm0.08$ mag for M87 from surface brightness fluctuation methods corresponding to a luminosity distance $D_L \sim 16.7\pm0.6$ Mpc \citep{blakeslee09}. After correcting the observed redshift $z_\mathrm{obs} = 0.004283$ for infall towards the Virgo cluster \citep[$z_\mathrm{corr}\sim 0.0032$;][]{mould00}, this $D_L$ translates to an angular size scale of 80.5 pc arcsec\per. We employed the optical velocity definition and referenced kinematic quantities in the barycentric frame by default.


\section{Data}
\label{sec:data_overview}

\subsection{Optical Imaging}
\label{sec:hst}

To create integration regions for this ALMA CO line search, we employed pipeline-calibrated HST images retrieved from the Mikulski Archive for Space Telescopes (DOI: \dataset[10.17909/wmbg-9051]{http://dx.doi.org/10.17909/wmbg-9051}). To isolate dust features, we used drizzled Wide Field Camera 3 \citep[WFC3/UVIS;][]{dressel22} F475W imaging from program GO-14256 (PI: Biretta). To map out filamentary H$\alpha$+[\ion{N}{2}] emission, we used a series of Advanced Camera for Surveys \citep[ACS;][]{ryon22} F660N polarized images from program GO-12271 (PI: Sparks) that we aligned before combining into a single Stokes $I$ image. In each case, we modeled and subtracted the stellar continuum using a concentric Multi-Gaussian Expansion \citep[MGE;][]{emsellem94} with uniform PAs using \texttt{GALFIT} \citep{peng02}. Following \citet{davidson24}, we iteratively expanded a pixel mask containing non-stellar features (gas and dust, the optical jet, and central AGN) to avoid contaminating the stellar fit. Figure~\ref{fig:hst} shows the final gas emission and dust absorption maps. In Section~\ref{sec:narrowem}, we describe the regions created for the CO emission-line search.

\subsection{Archival ALMA Imaging}
\label{sec:data_alma}

We obtained archival ALMA 12-m imaging from eleven single-pointing data sets taken between 2013 December 1 and 2018 September 25 whose phase centers coincided with the M87 nucleus (see Appendix~\ref{app:cont} and Table~\ref{tbl:sample} for additional details). In this nearly five-year window, one semipass program (2016.1.000415.S; PI: Marti-Vidal) was not included due to pipeline calibration difficulties arising from changing polarization calibrators between Execution Blocks \citep[EBs; for the continuum analyis, see][]{goddi21}. Another program (2013.1.00862.S; PI: Simionescu) was centered $\sim$40\arcsec\ south-east of the nucleus to image \cotwo\ in a dust and gas complex \citep[hereafter the SE cloud;][]{simionescu18}, and the field of view does not cover the M87 nucleus.

We also retrieved and analyzed ALMA Compact Array (ACA) 7-m observations in programs 2019.1.00807.S and 2021.1.01398.S. Due to large average FWHM of the synthesized beam $\overline{\theta}_\mathrm{FWHM} \gtrsim 5\arcsec$ and poorer limiting sensitivity compared to the ALMA 12-m data, these 7-m observations are only used for an analysis of the peak SED and to construct a light curve of the bright nucleus. We briefly discuss the ACA data sets, their calibration processes, and results in Appendix~\ref{app:specslope} and Table~\ref{tbl:aca_cont}.

\subsubsection{ALMA Data Properties}
\label{sec:data_prop}

The majority of the programs listed in Table~\ref{tbl:sample} were explicitly designed to probe nuclear continuum properties, although not all of these programs were obtained in time division mode (TDM). These heterogeneous data cover Bands 3, 4, 6, and 7 with $\overline{\theta}_\mathrm{FWHM}$ ranging from $\sim$0\farcs03 to 2\farcs3. The chosen spectral setups resulted in spectral resolutions from $\sim$1 to 63 MHz (after online Hanning smoothing). Both TDM-only and frequency division mode (FDM) data included dual and full polarization setups. For continuum-focused programs, a few redshifted atomic or molecular transitions (including low-$J$ CO) coincidentally lie within the spectral ranges, albeit at coarse spectral resolution ($\sim$20--100 \kms) at the native binning. Primary spectral windows (spws) for CO-line focused programs typically have velocity resolutions of $\sim$1--3 \kms. Calibrator observations of M87 were not included due to generally short integrations \citep[10 s to a few min; for further discussion, see][]{doi13}.

\begin{figure*}[!ht]
    \centering
    \includegraphics[width=\textwidth]{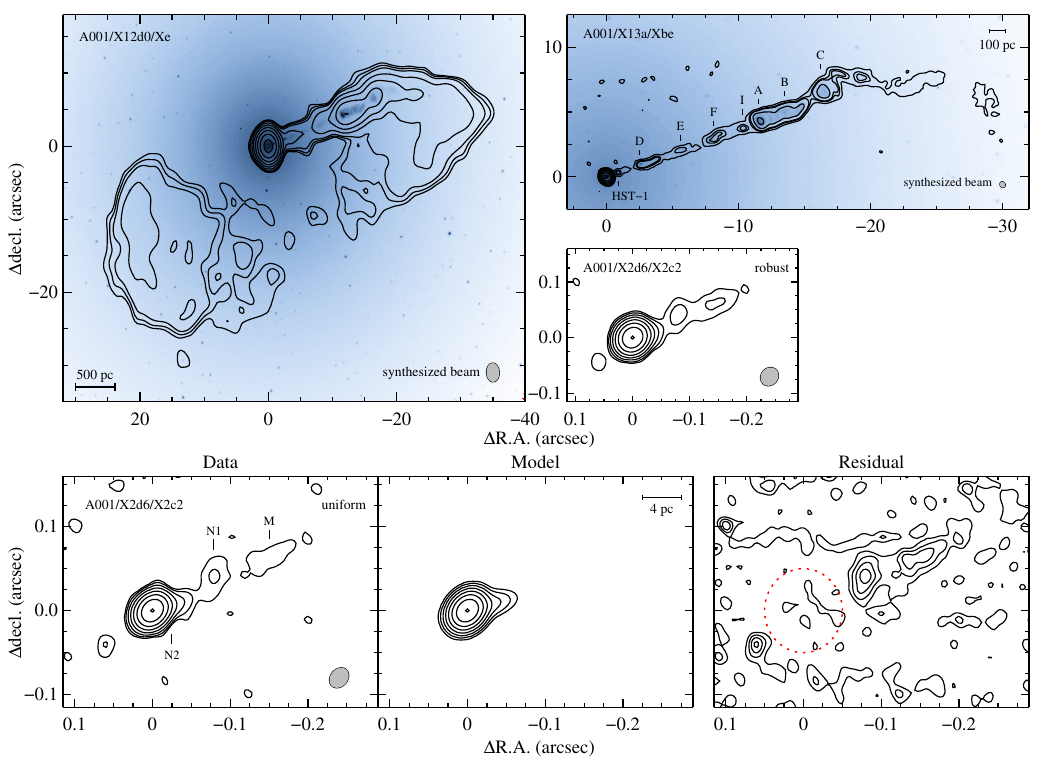}
    \caption{MFS imaging of M87 from ALMA data sets in Band 3 (contours; \textit{top panels}) and Band 6 (\textit{middle right}, \textit{bottom}) with the corresponding member OUS in each panel. For the two larger-scale panels, these mm-wavelength contours are overlain on an HST ACS F814W mosaic (from GO-13731; PI: Meyer). Some of these data sets are sensitive to the large-scale radio jet and lobes spanning a projected 5 kpc while others resolve the jet into individual clumps down to $\sim$2 pc from the core. The highest-resolution ALMA imaging is shown alongside a two-component \texttt{CASA} \texttt{imfit} model and residual map (\textit{bottom}; uniform weighting) that accounts for the central point source and an offset, semi-resolved Gaussian component with observed FWHM~$\sim 50 \times 19$ mas$^2$ to fit the blended N2 clump. Logarithmically-spaced continuum contours highlight high peak/rms dynamic range (with intensity contours reaching down to a DR~$\sim 2000$) while residual contours at 1, 3, 5, and 7$\times$rms show no additional continuum features in the central $R<0\farcs05$ ($\sim$4 pc; dashed circle) integration region. The excess feature to the southeast (at 7$\times$rms; rms = 0.83 mJy beam\per) is an imaging artifact and not part of the counter-jet.}
    \label{fig:mfs}
\end{figure*}

\subsubsection{Calibration Processes}
\label{sec:data_cal}

In general, we applied standard pipeline calibrations to the 12-m data using the appropriate \texttt{Common Astronomy Software Application} \citep[\texttt{CASA};][]{mcmullin07} pipeline version for each data set in order to construct Measurement Sets (MSs). Below, we discuss limited improvements to the standard pipeline calibration processes. Appendices~\ref{app:peak} and \ref{app:constraints} contain additional discussion of the bandpass calibration.

In adjacent or overlapping spws, peak continuum spectra show mismatches at the 1--3\% level, which is roughly consistent with relative flux calibration errors \citep{francis20b}. For projects 2015.1.00030.S and 2016.1.00021.S, these slight discrepancies complicated the $uv$-plane continuum subtraction for channel ranges with overlapping spws. Residual mismatches at the $\sim$0.1\% level impacted the search for very broad absorption or emission (see Figure~\ref{fig:contabs} and Appendix~\ref{app:broademission}). We manually re-scaled their flux calibrator \texttt{fluxdensity} Stokes $I$ levels in the \texttt{setjy} task for individual spws by up to 2\% to ensure more continuous flux densities in image-plane spectral cubes. 

We applied phase (and usually amplitude) self-calibration with final solution intervals that were per integration (per scan). In two cases in member OUSs A001/X12f/X20d and A002/X6444ba/X1b0, amplitude self-calibration was not adopted when the multi-frequency synthesis (MFS) image peak/rms dynamic range (DR) decreased noticeably compared to the final phase-only products. DRs spanned from several hundred up to $\sim$9000, far exceeding the more typical $\sim$100 without improved phase solutions \citep{richards22,privon24}. The highest $\mathrm{DRs} \gtrsim 5000$ were all obtained in Band 3. For certain other ALMA programs with bright continuum sources, higher DRs have been achieved with more careful, per-scan analysis \citep[e.g.,][]{komugi22}, especially when using brighter bandpass calibrators. However, those typically used for M87 are at the $>$2 Jy-level in Bands 3--7, effectively forestalling improved bandpass solution approaches \citep[e.g.,][]{oosterloo24}. The CO data cubes discussed in Section~\ref{sec:data_imaging} have spectral dynamical ranges (SDRs) that exceed $\sim$2000 per native (unbinned) channel in Band 3, although the bandpass stability limit imposes an effective SDR~$< 1000$ limit over large frequency ranges (see Appendix~\ref{app:bandpass}).

Three programs (2016.1.01154.V, 2017.1.00841.V, and 2017.1.00842.V) were obtained as a part of EHT+ALMA VLBI campaigns and most of their MSs coincidentally cover the redshifted \cotwo\ line. From the ALMA archive, the available .V datasets only contain columns and calibration tables pertaining to the phased (APP) observing mode. As is shown in Figure~\ref{fig:contabs}, APP-mode spectra show a characteristic scalloping pattern due to sets of 16 native channels (with individual binning $\Delta \nu_\mathrm{obs} = 7.81$ MHz) being used for coarse frequency-dependent phase solutions. We adopted the APP-mode results since other projects covered redshifted \cotwo\ and because reconstructing the ALMA interferometric-mode calibration tables is beyond the scope of this archival project \citep[for a guide, see][]{goddi19}. Without interferometric-mode calibration tables, phase self-calibration is not possible, and the peak (core) flux densities reported in Table~\ref{tbl:sample} are lower by about 30\% \citep{event21c,event24}. The scalloping spectral behavior prevents their inclusion in studies of broader line features ($\Delta v \gtrsim 50$ \kms), although some .V program data do reveal narrow absorption near redshifted $\nu_\mathrm{CO(2-1)}$.

\subsubsection{Manual Flagging}
\label{sec:data_flagging}

For dual polarization data obtained during ALMA Early Science, or for full polarization data when that was still a non-standard observing mode, calibration scripts from the archive often included manual flagging of the science target as well as the flux and bandpass calibrators. We manually flagged a limited number of additional visibilities based on high antenna temperature, amplitude disagreement, and poor bandpass calibration near spw edges. For the MS from member Observing Unit Set A001/X12f/X20f (hereafter X20f and likewise for other member OUS), the pipeline script manually flagged a moderately strong and variable atmospheric line in the bandpass calibrator as extragalactic in origin. This introduced a spurious and deep absorption feature in the peak continuum spectrum of M87 that was coincidentally very close to redshifted $\nu_\mathrm{CO(2-1)}$ \citep[see also][]{ray24}. Removing these manually-flagged channels, re-calibrating the raw visibilities, and re-imaging a new MS resulted in more consistent spectral behavior. Appendix~\ref{app:atm} contains more discussion about atmospheric line identification and variability.

\subsubsection{MFS and Spectral Imaging}
\label{sec:data_imaging}

As a part of the self-calibration loops, we imaged each MS into a Stokes $I$ MFS map using Briggs weighting \citep[with robust parameter $r=0.5$;][]{briggs95} in the \texttt{CASA} \texttt{tclean} deconvolution process. This weighting approach balanced sensitivity and angular resolution considerations. Some of these data were also deconvolved into MFS images using uniform (natural) weighting for greater resolution (sensitivity). Primary imaging results are given in Table~\ref{tbl:sample} while Figure~\ref{fig:mfs} shows MFS imaging for a representative sample of this archival project. Additional MFS imaging details and a brief continuum analysis are found in Appendices~\ref{app:cont} and \ref{app:specslope}. A more detailed analysis of the nuclear and extended continuum properties is beyond the scope of this paper but is covered in recent ALMA studies of the inner accretion flow and jet behavior \citep[e.g.,][]{lu23} as well as core and jet polarization \citep[e.g.,][]{goddi21,event21b,event21a,event21c}. Recent radio interferometric and VLBI studies have revealed detailed jet motion and internal structure \citep[e.g.,][]{hada16,hada17b,hada17a,pasetto21}.

High spectral resolution ALMA imaging of AGN have also revealed both Galactic and extragalactic narrow molecular absorption lines \citep[with typical FWHM~$\lesssim 20$ \kms; e.g.,][]{klitsch19,rose19b,rose24}. To that end, we first imaged each spw (or set of overlapping spws) using \texttt{tclean} with Briggs ($r=0.5$) weighting at the native channel binning given in Table~\ref{tbl:sample}, resulting in 72 continuum-dominated spectral cubes with a common restoring beam for each spectral setup. Peak flux densities are plotted in Figure~\ref{fig:contabs} across all unflagged frequencies for the entire sample. Absolute flux calibration uncertainties for MFS images and spectral cubes is about 5\% in Band 3 and 10\% in Bands 6 and 7 \citep[][]{fomalont14,francis20b,privon24}. For most spws in the archival sample, channels are coarsely binned ($\Delta v \sim 20-100$ \kms). Only four projects (2012.1.00661.S, 2013.1.00073.S, 2015.1.00030.S, and 2016.1.00021.S) have sufficiently fine FDM channels ($\Delta \nu_\mathrm{obs} \sim 0.6-3$ \kms) to enable detection of narrow molecular lines.

Lastly, we subtracted the continuum for MSs containing CO transitions using the \texttt{uvcontsub} task (with \texttt{fitorder} = 1) with three distinct fitting regions. In the first and second cases, we excluded channels corresponding to either $|\vlos - \vsys| < 500$ \kms\ to probe more standard velocity ranges for extragalactic gas or $<$1000 \kms\ for higher-amplitude rotation or outflow scenarios. Given the truncated \cothree\ velocity coverage in program 2012.1.00661.S, for the second case we only excluded the range $-1000 < \vlos - \vsys < 400$ \kms\ when subtracting the Band 7 continuum. In a third case for \coone\ only, we excluded the ranges $100 < |\vlos - \vsys| < 2200$ \kms\ to test for possible broad emission-line features that may originate from a pc-scale molecular CND. In this final case, including channels where $|\vlos - \vsys| < 100$ \kms\ helped to anchor the continuum subtraction near \vsys\ where any integrated CO emission is often minimal \citep[e.g.,][]{boizelle17,smith21}. We imaged the continuum-subtracted data into spectral cubes at both the native frequency spacing and at more coarse binning ($\sim$20$-$40 \kms) to increase the signal-to-noise ratio (S/N). In Sections~\ref{sec:line} and \ref{sec:disc}, we explore relatively narrow emission and absorption-line results derived using the first \texttt{uvcontsub} case. In Appendix~\ref{app:constraints}, we explore the second and third \texttt{uvcontsub} cases that produce very faint (and likely spurious) spectral deficits or excesses near redshifted CO transitions.

\begin{figure*}[!ht]
    \centering
    \includegraphics[width=\textwidth]{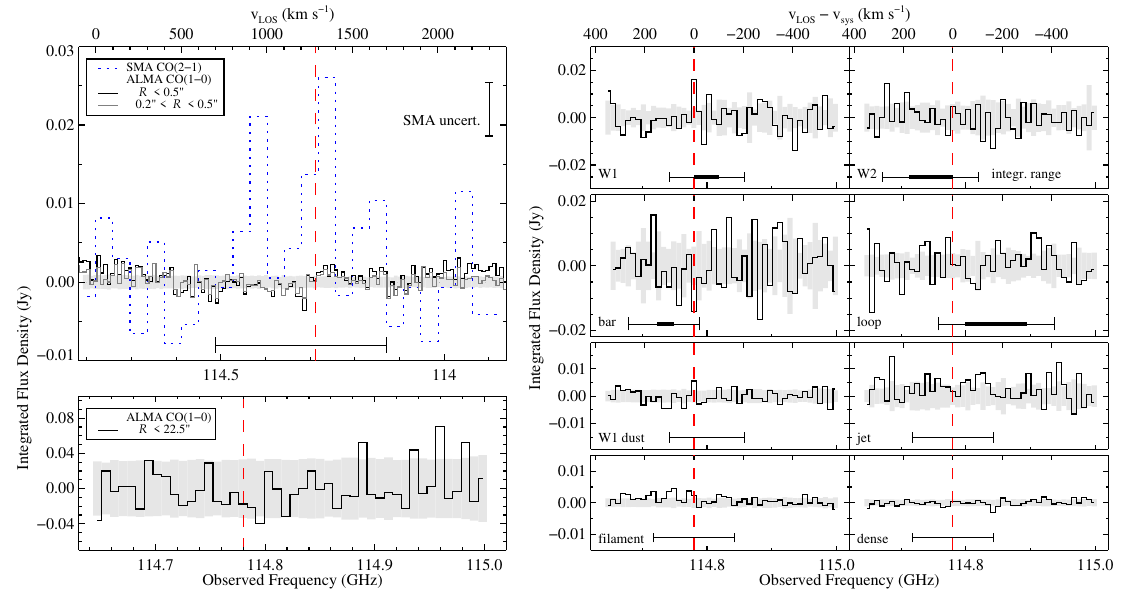}
    \caption{\coone\ line profiles and uncertainty ranges (shaded regions) extracted from the Xa1 continuum-subtracted spectral cube. Those centered on the nucleus include a circular aperture or annulus covering the CND extent with wide velocity extent (\textit{top left panel}) as well as a global \coone\ line profile out to a projected radius $R< 0.5\times$HPBW (\textit{bottom left}). In the nuclear spectrum with a circular aperture, a hint of central \coone\ absorption is still detected in the bin covering $\vlos - \vsys \sim -80$ \kms. The central SMA \cotwo\ profile from \citet[][]{tan08} was extracted using \texttt{PlotDigitizer}\footnote{\url{https://plotdigitizer.com}} and included for comparison. The remaining ALMA spectra (\textit{right panels}) were extracted using regions outlined in Figure~\ref{fig:hst}. Integration ranges (horizontal lines) are shown below the line profiles. For regions with filamentary or disk-like atomic gas emission, the approximate range of atomic gas velocities are included \citep[thick bars;][]{osorno23}. The redshifted $\nu_\mathrm{CO(1-0)}$ is shown in each panel assuming $z_\mathrm{obs} = 0.004283$ (dashed lines).}
    \label{fig:lineprof}
\end{figure*}

\section{Spectral Line Properties}
\label{sec:line}

Here, we focus on a spectral line search for extragalactic CO near the redshift of M87. ALMA spectra extracted against and around the M87 nucleus are remarkably featureless, confirming previous mm-wavelength studies in finding no strong extragalactic molecular lines in the inner $\sim$kpc \citep[e.g.,][]{braine93,tan08}. Integrated line profiles and peak continuum spectra in spws covering redshifted CO result primarily in $S_\mathrm{CO} \Delta v$ upper limits and low CO opacities. Additional dense gas tracers (e.g., SiO, CS, and HCN) are covered in certain data sets. Unfortunately, the expected low abundance of isotopologues or more exotic molecules relative to CO \citep{crocker12} and coarse spectral binning for those spws only allows for a CO line search. The ALMA 12-m spectra also show no evidence for Galactic \coone\ emission or absorption more directly about the M87 nucleus, even in spectra with binning $\Delta \nu_\mathrm{obs} < 3$ MHz that could resolve narrow Galactic lines \citep[whose line FWHM are often $\lesssim$10 \kms;][]{davies75,rugel18}.

A detailed mm/sub-mm line search is somewhat complicated by atmospheric lines that are not always fully removed during calibration \citep{shangguan20}. The residual impact is typically over a narrow frequency range and at the percent level or less, but it does affect extragalactic \cotwo\ detection for M87. In Appendix~\ref{app:atm}, we employ atmospheric transmission models to confirm the location and strength of telluric lines to avoid missing or misclassifying extragalactic lines.

\subsection{CO Emission-line Constraints}
\label{sec:narrowem}

For any detectable molecular gas in M87 accompanying the CND or individual filamentary features, we expected CO emission line speeds to be similar to those of the cospatial atomic gas. Within the CND, atomic gas speeds reach about $|\vlos - \vsys| < 500$ \kms. For filamentary structures out to a couple kpc, the gas speeds span $\pm$400 \kms\ \citep{boselli19,osorno23}. To conduct an efficient emission-line search, we extracted profiles from the X5 and Xa1 spectral cubes that had excluded channels with $|\vlos - \vsys| < 500$ \kms\ during continuum subtraction. These had the best combination of sensitivity and angular resolution to detect and isolate possible emission. Visibilities had been imaged with 20 \kms\ channels to better match the observed CO widths of other resolved cloud structures \citep[e.g., the SE cloud;][]{simionescu18}. We note, however, that no CO data cubes from other member OUS showed clear evidence for CO emission. \citet{ray24} highlight one channel in the X1b0 data cube that shows an apparently unresolved excess with a peak of 4.6 mJy beam\per\ (S/N~$\sim 7$). In Appendix~\ref{app:channel}, we demonstrate that this excess is likely spurious and not part of a larger-scale CND.

\begin{deluxetable*}{lccccccc}[ht]
\tabletypesize{\footnotesize}
\tablecaption{M87 \coone\ Emission-line Constraints\label{tbl:narrowem}}
\tablewidth{0pt}
\tablehead{
\colhead{Region} & \colhead{Area} & \colhead{$\overline{\mathrm{rms}}$} & \colhead{Peak $\nu_\mathrm{obs}$} & \colhead{Peak Excess} & \colhead{$\nu_\mathrm{obs}$ Range} & \colhead{min, max($\vlos - \vsys$)} & \colhead{Range Excess} \\[-1.5ex]
\colhead{Name} & \colhead{($N_\mathrm{beam}$)} & \colhead{(mJy)} & \colhead{(GHz)} & \colhead{(Jy \kms)} & \colhead{(GHz)} & \colhead{(\kms)} & \colhead{(Jy \kms)} \\[-1.5ex]
\colhead{(1)} & \colhead{(2)} & \colhead{(3)} & \colhead{(4)} & \colhead{(5)} & \colhead{(6)} & \colhead{(7)} & \colhead{(8)}
}
\startdata
    \multicolumn{8}{c}{\textbf{Atomic Gas Regions}} \\ \cline{1-8}
    W1 & 433.5 & 4.85 & 114.780 & 0.380 (0.113) & $114.741 - 114.858$ & $-205.2 , + 102.1$ & $<$0.414 \\
    W2 & 515.2 & 5.60 & 114.678 & 0.295 (0.109) & $114.670 - 114.819$ & $-102.8 , + 286.8$ & 0.639 (0.519) \\
    bar & 595.5 & 7.57 & 114.834 & 0.383 (0.129) & $114.678 - 114.788$ & $-20.9 , + 266.3$ & $<$0.633 \\
    loop & 253.3 & 4.46 & 114.741 & 0.230 (0.073) & $114.756 - 114.889$ & $-409.7 , + 61.1$ & 0.883 (0.464) \\
    jet & 150.0 & 4.23 & 114.827 & 0.276 (0.105) & $114.741 - 114.858$ & $-205.2 , + 102.1$ & 1.193 (0.355) \\\cline{1-8}
    \multicolumn{8}{c}{\textbf{Dust Regions}} \\ \cline{1-8}
    W1 dust & 45.9 & 2.32 & \nodata & \nodata & $114.741 - 114.858$ & $-205.2 , + 102.1$ & $<$0.191 \\
    filament & 20.9 & 1.44 & \nodata & \nodata & $114.717 - 114.842$ & $-164.3 , + 163.6$ & 0.502 (0.123) \\
    dense  & 10.1 & 1.00 & \nodata & \nodata & $114.717 - 114.842$ & $-164.3 , + 163.6$ & $<$0.085 \\\cline{1-8}
    \multicolumn{8}{c}{\textbf{Nuclear Regions}} \\ \cline{1-8}
    nucleus & 12.9 & 0.69 & \nodata & \nodata & $114.585 - 114.976$ & $-514.0 , + 510.6$ & $<$0.109 \\
    annulus & 10.8 & 0.68 & \nodata & \nodata & $114.585 - 114.976$ & $-514.0 , + 510.6$ & $<$0.085 \\
\enddata
\begin{singlespace}
  \tablecomments{\coone\ emission upper limits near the center of M87 from the Xa1 spectral cube with 20 \kms\ binning and typical rms~$\sim 0.4$ mJy beam\per\ per channel. Regions listed in col.\ (1) are shown in Figure~\ref{fig:hst} with corresponding beam-unit areas in col.\ (2). Col. (3) is the rms within each integration region averaged over the entire line profile. Cols.\ (4) and (5) give frequencies and integrated fluxes for the single channel with the highest flux density (if above S/N~$\gtrsim 3$) in each region. Cols.\ (6) and (7) give the channel frequency ranges and velocity endpoints corresponding to $\pm$100 \kms\ from the observed spread of atomic gas \vlos\ in each region or $\pm$150 \kms\ about \vsys\ for the dust-dominated regions. Col.\ (8) reports the possible CO fluxes (or upper limits) in these regions. Note that the peak excess in the integrated CO profile in Figure~\ref{fig:lineprof} does not always fall within the velocity integration ranges. Estimated 1$\sigma$ uncertainties (parentheses) were calculated by Monte Carlo resampling of the background noise using the same region shapes.}
\end{singlespace}
\end{deluxetable*}

In Figure~\ref{fig:hst}, we trace broad regions of diffuse ionized atomic gas emission and higher H$\alpha$/H$\beta$ decrement \citep[W1, W2, bar, and loop regions;][]{sparks04,osorno23} as well as more compact regions that cover higher dust opacities \citep[south of the nucleus and within region W1;][]{taylor20}. A final region in the jet (covering clumps A, B, and C; Figure~\ref{fig:mfs}) probes molecular gas in the shocked, higher-pressure ISM \citep{bicknell96}. From the F475W residual image, the dense dust region has an estimated intrinsic (deconvolved) maximum length of $\lesssim$65 pc; without the filamentary structure connecting on the east, however, the deepest absorption is more consistent with a width of $\sim$20 pc. The neighboring dusty filament and those in the W1 and loop regions appear to have intrinsic diameters of up to a few pc. Another dusty filament not covered in these apertures appears to span from the dense cloud to the west side of the loop region and comes close to the nucleus \citep[$R\lesssim0\farcs5$ separation; see Figure 6 from][and Figure~\ref{fig:hst} here]{event21c}.

In Figure~\ref{fig:lineprof} and Table~\ref{tbl:narrowem}, we present integrated \coone\ line profiles and fluxes for the intermediate-resolution Xa1 data set ($\overline{\theta}_\mathrm{FWHM} \sim 0\farcs 21$).  These line profiles reveal no unambiguous CO emission in any region, and the higher-resolution X5 data yields $\sim$2$\times$ poorer constraints due to the larger number of beams in each integration area. In regions where there is cospatial atomic gas seen in the F660N filter, we integrated line profiles over the range of atomic gas \vlos\ observed in (or around) these regions with an additional $\pm$100 \kms\ buffer. In dust-dominated regions seen primarily in the F475W filter, line profiles were integrated over a $\sim$300 \kms\ range centered on \vsys. To encompass the entire atomic gas CND, we integrated over a circular aperture with a radius $R = 0\farcs 5$ ($\sim$40 pc). An additional annulus integration between $0\farcs 2<R<0\farcs 5$ avoided larger nuclear residuals while still covering most of the ionized gas emission observed by \citet{osorno23}.

To estimate uncertainties, we followed a Monte Carlo (MC) resampling technique that iteratively shifted each region by a random amount before re-integrating \citep[without overlapping any defined regions;][]{boizelle17}. After 200 iterations, we adopted the standard deviation of these ``blank'' flux densities in each channel as a good proxy for the true uncertainty spectrum \citep[see also][]{tsukui23}. In some cases, the spread in ``blank'' flux densities is slightly higher (by at most 10\%) due to the impact of the primary beam correction further from the ALMA phase center, but this does not materially affect the analysis.

Based on the estimated noise spectra, the highest S/N candidate for CO emission is in the outer jet feature with an integrated $S_\mathrm{CO(1-0)}\Delta v \approx 1.19 \pm 0.36$ Jy \kms\ that just exceeds the 3$\sigma$ threshold. However, a channel-by-channel inspection of the integration over the velocity range shows no clear features. The dust filament and loop regions show positive features near \vsys\ at $2\sigma-3\sigma$ confidence, but the bulk of the signal being contained in 1$-$2 channels is not altogether convincing. Channel inspection did not reveal any likely discrete sources at those corresponding velocities (or elsewhere). A global \coone\ line profile integrated over the inner $\sim$2 kpc returns only an $S_\mathrm{CO(1-0)}\Delta v < 4.8$ Jy \kms\ 1$\sigma$ upper limit over the entire $\pm$500 \kms\ range.

The nuclear and annulus spectra also show no support for \coone\ emission. From typical rms uncertainties of 0.68 mJy beam\per\ in a 20 \kms\ channel, we estimated nuclear $S_\mathrm{CO(1-0)} \Delta v < 0.109$ Jy \kms\ (1$\sigma$ uncertainty) in a broad $|\vlos-\vsys| < 500$ \kms\ range to match the extraction region and velocity range of the previous SMA \cotwo\ study \citep{tan08}. The annulus region spectrum is better behaved than the composite nuclear spectrum, which shows excesses at large velocity offsets ($|\vlos - \vsys| \gtrsim 800$ \kms) due to higher continuum residuals directly at the nucleus location. In Appendix~\ref{app:constraints}, we discuss the anomalous spectral behavior when excluding an even broader channel range in the \texttt{uvcontsub} process.


\subsection{CO Absorption-line Properties}
\label{sec:narrowcoprops}

\begin{figure*}[!ht]
    \centering
    \includegraphics[width=0.49\textwidth]{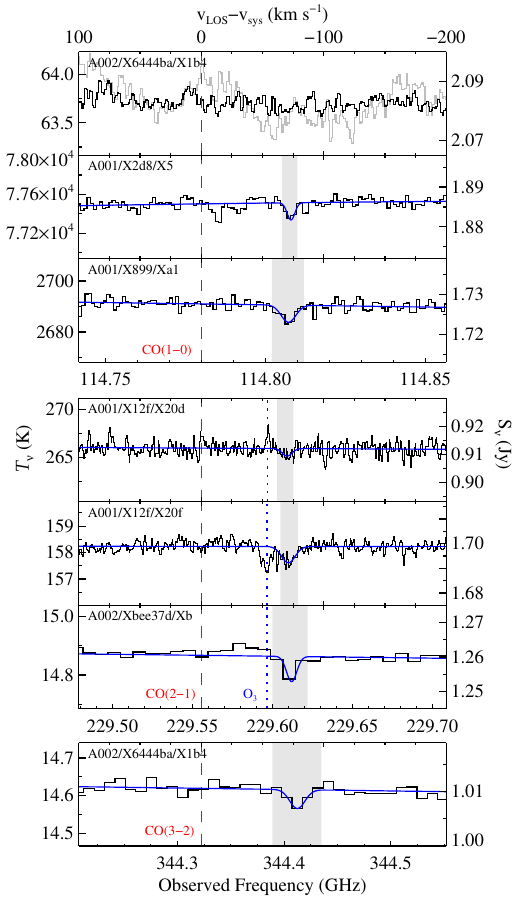}
    \includegraphics[width=0.49\textwidth]{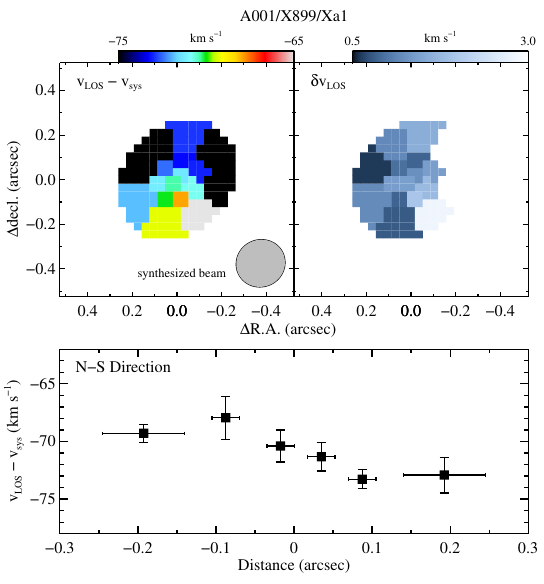}
    \caption{CO absorption lines detected against the bright nucleus of M87. Taken together, the peak continuum spectra at the native channel binning (\textit{left panels}; grouped by CO transition) provide unambiguous support for narrow, blueshifted CO absorption near the M87 redshift ($z_\mathrm{obs} = 0.004283$; red dashed lines). Only the X1b0 data set does not show this CO absorption-line feature. Its first EB spectrum (light gray) is very noisy while the higher-S/N second EB (black) still does not show faint \coone with peak opacity $\tau_0 \lesssim 0.003$ (or a depth of $S_\mathrm{\nu,cont}-S_0 \sim 0.006$ Jy). Gaussian fits (blue solid lines) show centroids between $-75 < \vlos - \vsys < -84$ \kms\ with narrow intrinsic line widths ($\sim$4$-$6 \kms). For the \cotwo\ transition, an atmospheric ozone line nearly overlaps with the (shaded) regions used to calculate integrated opacities. Spatial binning and Gaussian refitting of spectra extracted from the Xa1 continuum cube allowed for determination of $\vlos - \vsys$ values and velocity uncertainties $\delta \vlos$ across $\sim$2 beam widths (\textit{top right}; $\overline{\theta}_\mathrm{FWHM} \sim 0\farcs 21$). This limits a possible velocity gradient to be $\lesssim$3 \kms\ across $\sim$10 pc (\textit{bottom right}).}
    \label{fig:colineabsfit}
\end{figure*}

Despite the lack of any CO detection in emission, close inspection of the peak continuum spectra across 6 data sets in Figure~\ref{fig:colineabsfit} reveals blueshifted \coone, \cotwo, and \cothree\ absorption features between $-84 < \vlos - \vsys < -75$ \kms. Only one CO spw with narrow $\Delta \nu_\mathrm{obs} \lesssim 1$ MHz binning from the X1b0 data set returns a non-detection, and its more complicated spectral response shown in Figure~\ref{fig:contabs} is ameliorated by using just the lowest-rms data from the second EB\footnote{\citet{ray24} use both EBs and find a complex response that is qualitatively very similar to the peak spectrum in Figure~\ref{fig:contabs}. Their fitted absorption-line components have $\vlos-\vsys = -527$ \kms\ and $-$74 \kms\ with $\sigma \sim 230$ \kms\ and $\sim$46 \kms, respectively.} However, this more selective spectrum still appears featureless (see Figure~\ref{fig:colineabsfit}) due to a sensitivity limit that exceeds the expected opacity spectrum in the common definition:

\begin{equation}
    \tau_\nu = -\ln \left(\frac{T_\nu}{T_{\nu,\mathrm{cont}}}\right)\,.
\end{equation}

\noindent Here, the peak continuum spectrum $T_\nu \propto S_\nu/\theta_\mathrm{maj}\theta_\mathrm{min}$ is measured in K units and $T_{\nu,\mathrm{cont}}$ is the fitted linear continuum level about the CO line feature. For the most confident cases, the measured peak $\tau_0 \lesssim 0.004$ values are much lower than for CO absorption lines observed in other ETGs \citep[typically $\tau_0 \sim 0.1 - 0.2$, but may approach unity;][]{boizelle17,ruffa19,klitsch19,rose19b,rose19a,rose20,rose24,kameno20}. No other potential extragalactic CO or dense gas absorption lines are seen at a consistent $\vlos-\vsys$ across multiple data sets or transitions, including at the same $\vlos-\vsys \sim -131$ \kms\ for the atomic-gas absorber.

From these opacity spectra, we also calculated upper limits of the integrated line opacity

\begin{equation}
    \tau_\mathrm{CO} = \int \tau_\nu dv
\end{equation}

\noindent for CO absorption in \kms\ units. Across these three transitions, the $\tau_\mathrm{CO} \sim 0.02-0.05$ \kms\ range (with a median $\mathrm{S/N} \sim 4$) is very low and better matches the bottom end of the $\tau_\mathrm{CO}$ distribution for Galactic diffuse molecular clouds \citep[e.g.,][]{liszt98,liszt12,liszt19}.

\cotwo\ absorption lines are detected less confidently than are those for \coone, with an extreme case from the X20d data that has just S/N~$\sim 1.4$ for its $\tau_\mathrm{0,CO(2-1)}$ measurement. The $\tau_\mathrm{CO(2-1)}$ values may be suppressed by $\sim$20$-$30\% due to difficulty in fully removing the neighboring ozone line contamination (see Appendix~\ref{app:atm}). Two APP-mode programs cover the redshifted $\nu_\mathrm{CO(2-1)}$, but these peak spectra display frequency-dependent fluctuations due to phasing all elements together into a single array. Some of these .V program spectra appear to show \cotwo\ absorption (see Figure~\ref{fig:contabs}), but we only utilize spectral fits to the most regular Xb data set from program 2016.1.01154.V when calculating CO absorber properties.

\begin{deluxetable*}{lccccccc}[ht]
\tabletypesize{\footnotesize}
\tablecaption{M87 Narrow CO Absorption-line Properties\label{tbl:narrowabs}}
\tablewidth{0pt}
\tablehead{
\colhead{Member} & \colhead{rms} & \colhead{$\nu_\mathrm{LOS}$} & \colhead{$\vlos-\vsys$} & \multicolumn{2}{c}{\sigmalos} & \colhead{$\tau_0$} & \colhead{$\tau_\mathrm{CO}$} \\[-1.5ex]
\colhead{OUS} & \colhead{(mJy beam\per)} & \colhead{(GHz)} & \colhead{(\kms)} & \colhead{(MHz)} & \colhead{(\kms)} &  & \colhead{(\kms)} \\[-1.5ex]
\colhead{(1)} & \colhead{(2)} & \colhead{(3)} & \colhead{(4)} & \colhead{(5)} & \colhead{(6)} & \colhead{(7)} & \colhead{(8)}
}
\startdata
    \multicolumn{8}{c}{\textbf{\coone}} \\ \cline{1-8}
    A002/X6444ba/X1b0 & 2.61 & \nodata & \nodata & \nodata & \nodata &  & $<0.056$ \\
    A001/X2d8/X5 & 0.91 & 114.808 (0.000287) & $-$78.45 (0.75) & 1.204 (0.317) & 3.13 (0.82) & 0.002286 (0.000785) & 0.0220 (0.0059) \\
    A001/X899/Xa1 & 1.15 & 114.807 (0.000468) & $-$76.38 (1.22) & 2.275 (0.526) & 5.92 (1.37) & 0.002483 (0.000735) & 0.0361 (0.0068) \\[1.5ex] \cline{1-8}
    \multicolumn{8}{c}{\textbf{\cotwo}} \\ \cline{1-8}
    A001/X12f/X20d & 2.38 & 229.609 (0.001600) & $-$75.29 (2.08) & 3.012 (1.977) & 3.92 (2.57) & 0.002989 (0.002185) & 0.0325 (0.0078) \\
    A001/X12f/X20f & 1.89 & 229.610 (0.000533) & $-$76.31 (0.69) & 4.259 (0.527) & 5.54 (0.69) & 0.003904 (0.000677) & 0.0438 (0.0053) \\
    A002/Xbee37d/Xb & 0.85 & 229.612 (0.001210) & $-$78.94 (1.57) & 3.211 (1.340) & 4.18 (1.73) & 0.005798 (0.003290) & 0.0461 (0.0109) \\[1.5ex] \cline{1-8}
    \multicolumn{8}{c}{\textbf{\cothree}} \\ \cline{1-8}
    A002/X6444ba/X1b4 & 0.95 & 344.412 (0.003276) & $-$83.58 (2.84) & 7.119 (3.592) & 6.17 (3.11) & 0.003418 (0.002263) & 0.0574 (0.0174) \\
\enddata
\begin{singlespace}
  \tablecomments{CO absorption-line properties measured from the M87 peak continuum spectra. Col.\ (2) gives the typical rms noise per channel. Cols.\ (3) to (6) are best-fit velocity and line width parameter values from Gaussian fitting with the $\sigmalos$ values being intrinsic widths before any online Hanning smoothing. Cols.\ (7) and (8) report the peak and integrated CO opacities after continuum fitting and normalization. Uncertainties in these fitted Gaussian parameters and opacities (in parentheses) were estimated using a Monte Carlo resampling technique.}
\end{singlespace}
\end{deluxetable*}

To better characterize these line properties, we fit each CO absorption feature with a Gaussian line profile and a linear continuum to recover the line centroid ($\nu_\mathrm{LOS}$, \vlos), line width (\sigmalos), and peak opacity ($\tau_0$). Results are presented in Figure~\ref{fig:colineabsfit} and Table~\ref{tbl:narrowabs}. These fits incorporated Hanning smoothing\footnote{In frequency space, Hanning smoothing operates as a convolution of the FDM data before on-line channel averaging. The triangular kernel consists of three channels with central and side amplitudes of 0.50 and 0.25, respectively. Finely-binned FDM data have frequency resolutions up to 2$\times$ larger due to Hanning smoothing. For those with the narrowest $\Delta\nu_\mathrm{obs} < 1$ MHz, the intrinsic line widths are $\sim$7\% lower than the blurred values. For those with $\Delta\nu_\mathrm{obs}>5$ MHz, there is no change in frequency resolution.} to recover intrinsic \sigmalos. We estimated parameter uncertainties by an MC resampling technique that adds random noise to the best-fitting model spectrum before refitting. The noise was drawn from a normal distribution with a standard deviation equal to that of line-free portions of each spectrum. During this process, we also explored channelization effects by randomly shifting the bin centers by a uniform fraction of a channel width when constructing the model spectrum. These frequency shifts were also drawn from a normal distribution with standard deviation equal to a quarter of the corresponding $\Delta \nu_\mathrm{obs}$ value in Table~\ref{tbl:sample}. This final step helps to provide more realistic line-width uncertainties in cases where the measured width $\sigma_\mathrm{LOS} \lesssim \Delta\nu_\mathrm{obs}$. Incorporating Hanning smoothing also better accounts for correlated spectral noise, although it cannot explain apparently coherent spectra features over frequency ranges $\gtrsim$3$\Delta \nu_\mathrm{obs}$ (e.g., see the X5 spectrum in Figure~\ref{fig:colineabsfit}). Broader uncertainties on $\tau_\mathrm{CO}$ were also determined in this way. We note that bandpass calibration limitations are liable to introduce pattern noise in the peak continuum spectrum over larger frequency ranges (at the $\sim$0.1\% level, or $\sim$1$-$2 mJy; see Figure~\ref{fig:contabs}). This may create low-level structure in the spectral response over the  model fitting region that is not represented in the linear $T_\mathrm{\nu,cont}$ fit. This appears most clearly in the \cothree\ spectrum (see Figure~\ref{fig:contabs}). However, we do not attempt to quantify here the bandpass calibration effects on $\tau_0$ and $\tau_\mathrm{CO}$ values and simply refer the reader to Appendix~\ref{app:broadabs} for additional discussion.

The ratio of centroid frequencies (or weighted-average $\overline{\nu}$ when multiple observations exist) is redshift independent, and the observed ratios are broadly consistent with expectations:

\begin{equation*}
    \begin{aligned}
        \frac{\nu_\mathrm{CO(3-2)}}{\overline{\nu}_\mathrm{CO(1-0)}} -3  & = (-8.96\pm 2.67) \times 10^{-5} \\
        \frac{\overline{\nu}_\mathrm{CO(2-1)}}{\overline{\nu}_\mathrm{CO(1-0)}} -2 & = (-4.15\pm 0.58)\times 10^{-5}\,.
    \end{aligned}
\end{equation*}

\noindent Deviations from the expected zero values beyond the standard error propagation may result from low-S/N \cotwo\ detection. We note that including the APP-mode value in Table~\ref{tbl:narrowabs} when calculating $\overline{\nu}_\mathrm{CO(2-1)}$ does not contribute appreciably to the measured $\overline{\nu}_\mathrm{CO(2-1)}/\overline{\nu}_\mathrm{CO(1-0)}-2$ offset from zero. As shown in Figure~\ref{fig:colineabsfit}, the faint \cotwo\ absorption and neighboring O$_3$ line would not have allowed for a confident CO detection without support from the \coone\ transition. For the other ratio, the $\nu_\mathrm{CO(3-2)}$ value appears to be anomalously high. In velocity units, these fits give consistent (intrinsic) line widths of $\sim$3$-$6 \kms\ and typical $\vlos - \vsys \sim -75$ to $-$79 \kms, although the \cothree\ feature has a more blueshifted central velocity that is discrepant by $\sim$3$\sigma$ from the weighted-average \coone\ and \cotwo\ centroid values.

For use later when constraining molecular gas properties, we also constructed final weighted-average peak opacities $\overline{\tau}_\mathrm{0,CO(1-0)} = 0.00239\pm 0.00054$ and $\overline{\tau}_\mathrm{0,CO(2-1)} = 0.00390\pm 0.00063$. If ignoring the APP-mode results from the Xb data set, $\overline{\tau}_\mathrm{0,CO(2-1)}$ only changes by $\sim$2\%. The integrated opacity $\overline{\tau}_\mathrm{CO(1-0)} = 0.0281 \pm 0.0045$ \kms\ is moderately lower than measured for $\tau_\mathrm{CO(3-2)}$ while the latter is formally consistent with $\overline{\tau}_\mathrm{CO(2-1)} = 0.0410 \pm 0.0041$ \kms. We note that these $\tau_\mathrm{CO}$ are an order-of-magnitude lower than those reported by \citet{ray24}.

The core continuum source at the center of M87 is very compact, with the brightest component having a radial size of $\sim$25 $\mu$as \citep[0.002 pc;][]{event19} that is expected to probe a very narrow line of sight through the CO absorber. At high resolution \citep[e.g.,][]{acciari09}, this very compact core brightness may account for only $\sim$25$-$50\% of the integrated flux density that remains unresolved in these ALMA data after extrapolating the flat-spectrum core and synchrotron jet behavior below 3.5 mm \citep[with spectral index $\alpha \sim -0.6$ for $S_\nu \propto \nu^\alpha$; Figure~\ref{fig:cont_sed}; see also][]{doi13,lu23}. As discussed in Appendix~\ref{app:mfs}, the next brightest compact synchrotron source (the N2 clump) lies $<$2 pc in projection from the nucleus with an integrated mm-wavelength core-to-N2 flux density ratio of $\sim$25. On smaller scales at radio wavelengths, the synchrotron core-to-jet flux density ratio approaches $\sim$100 \citep[e.g.,][]{cheung07,walker18}.

The question is whether the continuum emission from the approaching synchrotron jet emission (oriented at a $\mathrm{PA}\sim -65\degr$ from the core) passes through sufficiently distinct lines of sight to probe a possible absorption-line velocity gradient in the CO absorber. We attempted to answer this by refitting the \coone\ absorption feature from the Xa1 data set across the central beam area ($\overline{\theta}_\mathrm{FWHM} \sim 0\farcs 21$ or $\sim$15 pc). Spectra in adjacent spatial pixels were combined together using Voronoi tessellation \citep{cappellari03} prior to refitting to achieve roughly uniform continuum S/N per bin. Figure~\ref{fig:colineabsfit} shows the velocity gradient to be no more than 3 \kms\ across $\sim$10 pc, with the putative gradient being oriented in a mostly N--S direction. While not unphysical for emission-line clouds in luminous ETGs \citep[e.g.,][]{utomo15}, residual calibration issues erode support for a rotating cloud scenario (for additional details, see Appendix~\ref{app:broademission}). Individual uncertainties were estimated using the same MC resampling approach. The other \coone\ continuum cube from the X5 data set had much higher angular resolution but insufficient S/N in binned spectra to measure opacities across the central beam area.

\begin{figure*}
    \centering
    \includegraphics[width=\textwidth]{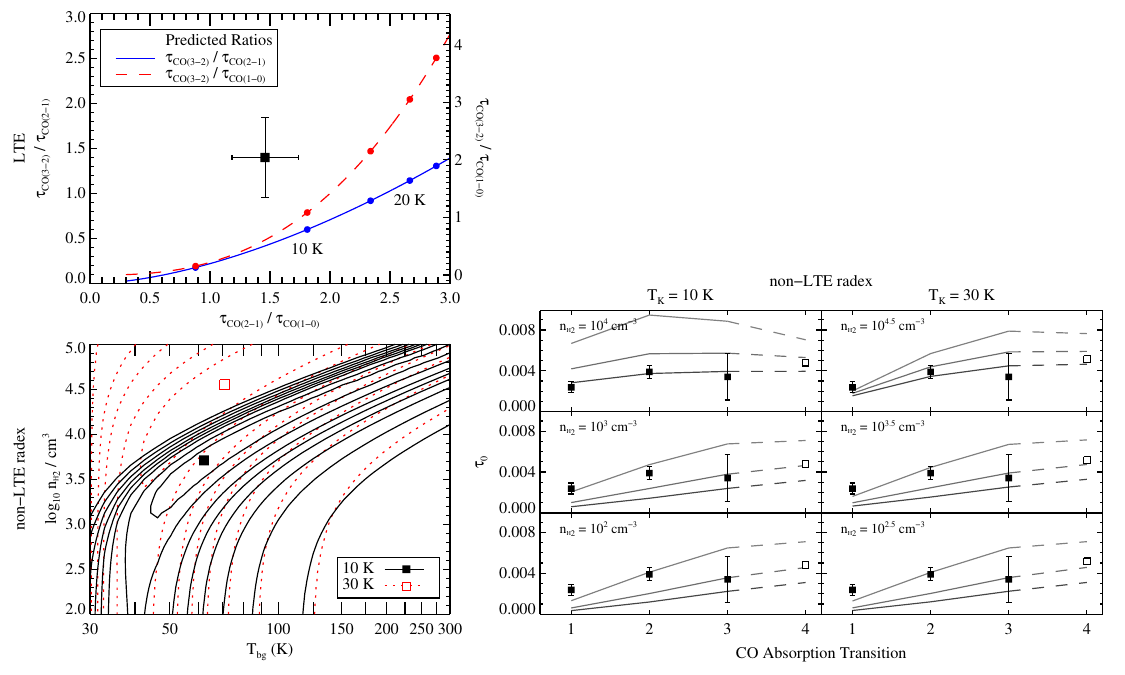}
    \caption{Comparison between observed and predicted absorption-line opacities for the M87 CO absorber. Integrated opacity ratios (\textit{top left panel}) do not agree with LTE predictions (circles, showing $T_\mathrm{ex}$ separated by 5 K). Non-LTE \texttt{radex} modeling \citep{van07} for cold gas ($T_K \leq 30$ K) using the measured absorption-line FWHM~$\approx 10$ \kms\ and $N_\mathrm{CO} \approx 10^{15}$ cm\pertwo\ better reproduces the observed peak opacities (\textit{bottom left}, \textit{right}). Best-fit \texttt{radex} results for $T_K = 10$ K have $n_\mathrm{H_2} \sim 5000$ cm\perthree\ embedded in a hotter background temperature $T_\mathrm{bg} \sim 60$ K. Inner $\Delta\chi^2 = \chi^2 - \min(\chi^2)$ contours are at the 1, 2, and 3$\sigma$ level while the remainder are arbitrarily (linearly) spaced. For comparison, the measured and \texttt{radex} model $\tau_0$ are shown (\textit{right}) for a range of input $n_\mathrm{H_2}$, $T_\mathrm{bg}$, and $T_K$ parameters together with predicted $\tau_\mathrm{0,CO(4-3)}$ values (open squares) for the best-fitting parameters for each $T_K$ scenario (\textit{bottom left}).}
    \label{fig:tex_plot}
\end{figure*}

\subsection{Temperature and Column Densities}
\label{sec:Tex}

The relative level occupation defines an excitation temperature $T_\mathrm{ex}$ for two different transitions. Given the higher confidence in the lower-$J$ $\tau_\mathrm{CO}$, we first solved column density equations that relate integrated \cotwo\ and \coone\ opacities with a common $T^{21}_\mathrm{ex}$ \citep{godard2010,rose19b} by

\begin{equation}
    \frac{\tau_\mathrm{CO(2-1)}}{\tau_\mathrm{CO(1-0)}} \left[\frac{\nu_\mathrm{CO(2-1)}}{\nu_\mathrm{CO(1-0)}}\right]^3 \frac{g_{1}A_{10}}{g_{2}A_{21}} = \frac{1-e^{-h\nu_\mathrm{CO(2-1)}/kT^{21}_\mathrm{ex}}}{e^{h\nu_\mathrm{CO(1-0)}/kT^{21}_\mathrm{ex}}-1} \,.
\end{equation}

\noindent A similar equation relates a common $T^{32}_\mathrm{ex}$ for adjacent \cothree\ and \cotwo\ transitions. Here, $g_u = 2 J_u +1$ is the upper-level degeneracy for that radiative transition and $A_{ul}$ is the corresponding Einstein coefficient for spontaneous decay\footnote{Obtained from Splatalogue: \url{https://splatalogue.online/\#/advanced}}. This assumes a single-zone system in local thermodynamic equilibrium (LTE) where adjacent transitions (e.g., $J = 0\rightarrow1$ and $1\rightarrow2$) originate under the same conditions. We have ignored beam filling factors and also considered the impact of the CMB background to be negligible at this redshift and for the measured $T^{21}_\mathrm{ex}$ and $T^{32}_\mathrm{ex}$ values. Individual $\tau_\mathrm{CO}$ have lower S/N, so we employed weighted-average values to recover more confident $T_\mathrm{ex}$.

In Figure~\ref{fig:tex_plot}, we compare measured opacity ratios with theoretical values assuming the absorbing medium is in LTE with $T^{32}_\mathrm{ex} = T^{21}_\mathrm{ex}$. However, the relatively large depth of the \cothree\ feature and/or low \cotwo\ depth as evidenced by $\tau_\mathrm{CO(3-2)}/\overline{\tau}_\mathrm{CO(2-1)} \approx \overline{\tau}_\mathrm{CO(2-1)}/\overline{\tau}_\mathrm{CO(1-0)}$ is not consistent with LTE expectations. This may indicate an elevated $J=2$ population, possibly arising from absorption-line contributions by gas in distinct spatial or energetic regions \citep[e.g., cloud outskirts; see][]{rose24}. Unfortunately, neither $\tau_\mathrm{CO(3-2)}$ or $\overline{\tau}_\mathrm{CO(2-1)}$ values are as reliable as is $\overline{\tau}_\mathrm{CO(1-0)}$, and we do not attempt any reconciliation in the LTE framework. In Section~\ref{sec:disc_temp}, we use radiative transfer modeling to better account for possible deviations from LTE.

To determine a total column density $N_\mathrm{CO}$, we primarily considered \coone\ absorption and an excitation temperature for the lowest-$J$ transitions:

\begin{equation}
    N_\mathrm{CO} = Q_\mathrm{rot} \left(\frac{8 \pi v_{ul}^3}{c^3}\right) \frac{g_0}{g_1} \frac{1}{A_{10}} \frac{1}{1-e^{-h\nu_\mathrm{CO(1-0)}/kT^{21}_\mathrm{ex}}} \tau_{\mathrm{CO(1-0)}} \,.
\end{equation}

\noindent This relationship assumes detailed balancing and the temperature-dependent $^{12}$CO rotational partition function $Q_\mathrm{rot}$ \citep{Mangum2015}. Assuming LTE, the kinetic temperature $T_K = T_\mathrm{ex}$. Here, we used $\overline{\tau}_\mathrm{CO(1-0)}$ for more confident $N_\mathrm{CO}$ determination.

We recovered more reliable error bars in a series of trials by drawing individual $\tau_\mathrm{CO}$ randomly from normal distributions centered on their measured value with standard deviations equal to their 1$\sigma$ uncertainties in Table~\ref{tbl:narrowabs}. After calculating weighted-average opacities at each iteration, we used the newly-fit $T_\mathrm{ex}$ to find a new $Q_\mathrm{rot}$ and $N_\mathrm{CO}$. After $10^5$ iterations, we used the 15.9 to 84.1 (0.14 to 99.87) percentile levels of the $T^{21}_\mathrm{ex}$ distribution to find a median $T^{21}_\mathrm{ex} = 7.77^{+1.34}_{-1.94}\left(^{+14.88}_{-2.90}\right)$ K and $N_\mathrm{CO}/10^{15}\,\mathrm{cm}\pertwo = 1.16^{+0.21}_{-0.28}\left(^{+3.55}_{-0.47}\right)$ with 1$\sigma$ (3$\sigma$) error bars. Calculated separately for the higher-$J$ lines, we find higher median $T^{32}_\mathrm{ex} = 28.6^{+26.7}_{-14.9}$ K with $N_\mathrm{CO} \sim 10^{16}$ cm\pertwo\ whose high 1$\sigma$ error bars arise from relatively large $\tau_\mathrm{CO(3-2)}$ uncertainty. Coincidentally, \citet{ray24} report similar $T_\mathrm{ex}^{21}$ while their $N_\mathrm{CO}$ is $\sim$5$\times$ lower using the X1b0 data set.

\section{Discussion}
\label{sec:disc}

\subsection{CO Emission}
\label{sec:disc_em}

Based on line profile upper limits in Table~\ref{tbl:narrowem}, these archival ALMA data do not support any CO emission-line detection from dusty clumps or atomic gas filaments. Integrated line profiles for the larger-scale W1, W2, bar, and loop regions have individual detection thresholds at the $\sim$0.4$-$0.6 Jy \kms\ level. For a standard \coone-to-H$_2$ conversion factor $\alpha_\mathrm{CO} = 3.1$ \msun\ pc\pertwo\ (K \kms)\per\ \citep{Sandstrom2013}, this translates to $M_\mathrm{H_2} \lesssim (0.8-1.3)\times 10^6$ \msun\ upper limits. In the nucleus ($R<0\farcs 5$), we find a limiting $S_\mathrm{CO(1-0)} \Delta v < 0.109$ Jy \kms\ with a 20\% lower value for the annulus region. The resulting $M_\mathrm{H_2} < 2.3\times 10^5$ \msun\ is $\sim$20$\times$ lower than any previous upper limit \citep{tan08}. Both this $M_\mathrm{H_2}$ limit and the corresponding (deprojected) surface mass density $\overline{\Sigma}_\mathrm{H_2} < 45$ \msun\ pc\pertwo\ averaged over the atomic CND extent are lower than for many CO-detected molecular CNDs in other massive elliptical galaxies \citep{boizelle17,ruffa19,smith21}. Such a tight $M_\mathrm{H_2}$ constraint is in stark contrast to the estimated $\sim$10$^7$ \msun\ within the inner $\sim$200 pc reported by \citet{ray24}.

The single-channel ($\Delta v = 20$ \kms) CO line profile excesses in the W1 and bar regions and an integrated excess ($\Delta v = 100$ \kms) in the jet region over clumps A--C \textit{may} hint at CO emission. However, the low estimated S/N~$\sim 3$ and a lack of clear compact or diffuse CO emission at the corresponding locations and frequencies disfavor that interpretation. For broad detection, the only possible candidate is the dust filament found $\sim$2\arcsec\ to the southeast of the nucleus, although its integrated line profile only reaches S/N~$\sim 4$ in the $0 < \vlos - \vsys < 320$ \kms\ range. Again, a channel-by-channel inspection of this region does not show any potential features above a 4$\times$rms threshold.

Unfortunately, new ALMA 12-m CO imaging would likely not result in \coone\ detection from either the known dusty regions or the CND. To achieve just a 3$\times$ improvement in noise over the Xa1 results (reaching an rms~$\sim 0.225$ mJy beam\per\ in a 20 \kms\ channel), the ALMA \texttt{Observing Tool} \cite[\texttt{OT};][]{immer24} suggests a needed on-source (total) time of 8 hrs (12.5 hrs). Further improvement with SDR~$\gtrsim 3000$ in a 20 \kms\ channel is likely beyond the current ALMA Band 3 capabilities. Unless the CND has an $M_\mathrm{H_2} \sim (0.75-2)\times 10^5$ \msun, future ALMA CO line surveys of the M87 nucleus are not likely to detect any molecular gas in emission.

HST imaging of the dense and filament regions adjacent to the nucleus in projection provide some additional constraints on possible CO emission. Assuming close proximity to the M87 midplane, this dust should obscure $\sim$half of the stellar light along those lines of sight \citep[for more details, see][]{ferrarese96,viaene17,boizelle19}. In this scenario, observed fractional deficits in the F475W data (see Figure~\ref{fig:hst}) correspond to peak \textit{intrinsic} $A_\mathrm{F475W} = 0.75$ mag and 0.25$-$0.4 mag for the dense and filamentary regions, respectively. Since the M87 dust-to-gas ratio is estimated to be close to MW values \citep{sparks93}, an expected $A_V / A_\mathrm{F475W} \sim 0.84$ \citep{cardelli89} and standard dust-to-extinction correlations \citep{draine14} return a peak dust surface mass density $\Sigma_\mathrm{dust} \sim 0.1$ \msun\ pc\pertwo\ and suggest an average of $\sim$0.05 \msun\ pc\pertwo\ over the inner portion of the dense cloud region. Taking an intrinsic (deconvolved) radius ($\lesssim$20 pc; see Section~\ref{sec:narrowem}) for the obscuration as a proxy for the cloud size and a standard dust-to-gas mass of $\sim$0.015 \citep{Sandstrom2013}, the expected $\overline{\Sigma}_\mathrm{H_2} \lesssim 3$ \msun\ pc\pertwo\ and a total $M_\mathrm{H_2} \lesssim 4200$ \msun\ is consistent with ALMA CO non-detection. Assuming the same $\alpha_\mathrm{CO}$ value, the corresponding integrated $S_\mathrm{CO(1-0)}\Delta v$ would be about 75$\times$ lower than the upper limit found for the dense dust aperture.

Cool-core BCGs frequently host rotation-dominated CNDs and/or extended filaments with CO-derived total molecular gas masses $M_\mathrm{H_2} \sim 10^8$ \msun\ \citep{Olivares2019,Baek2022}, and CO surveys of galaxies like M87 have found a positive correlation between integrated H$\alpha$ luminosity $L_\mathrm{H\alpha}$ and total $M_\mathrm{H_2}$ \citep[e.g.,][]{edge01,salome03,pulido18}. Among these massive elliptical galaxies, M87 has a low but confident\footnote{Not all cool-core galaxies host CNDs that could contribute to the global $L_\mathrm{H\alpha}$. The atomic CND in M87 contributes 20$-$25\% of its global $L_\mathrm{H\alpha}$  \citep{sparks93,walsh13,osorno23}.} $L_\mathrm{H\alpha}\sim (0.8-1.1)\times 10^{40}$ erg s\per, and the $L_\mathrm{H\alpha} - M_\mathrm{H_2}$ correlation suggests there should be a global $M_\mathrm{H_2} \sim (0.8-9)\times 10^8$ \msun. Previous low-$J$ CO observations of M87 found a global $M_\mathrm{H_2}<10^8$ \msun\ \citep{jaffe87}. Adding together all the emission-line search regions in Figure~\ref{fig:hst}, we place a more strict $M_\mathrm{H_2} \lesssim 3\times 10^6$ \msun. When using a circular aperture with 2 kpc radius, we find a similar $<$4$\times$10$^6$ \msun\ global upper limit. This ALMA CO search suggests that M87 is either a clear outlier in the $L_\mathrm{H\alpha} - M_\mathrm{H_2}$ correlation for cool-core galaxies or that there is much wider scatter about this correlation.

\subsection{CO Absorption}
\label{sec:disc_abs}

Here, we discuss the properties of extragalactic CO absorption seen against the M87 nucleus. Based on both CO opacities and kinematics, this CO absorption appears more consistent with a narrow filament instead of a gravitationally bound cloud. We place the observed kinematics, density estimates, and temperatures in context with circumnuclear gas seen both in absorption and emission in other luminous ETGs.

\begin{figure}
    \centering
    \includegraphics[width=\columnwidth]{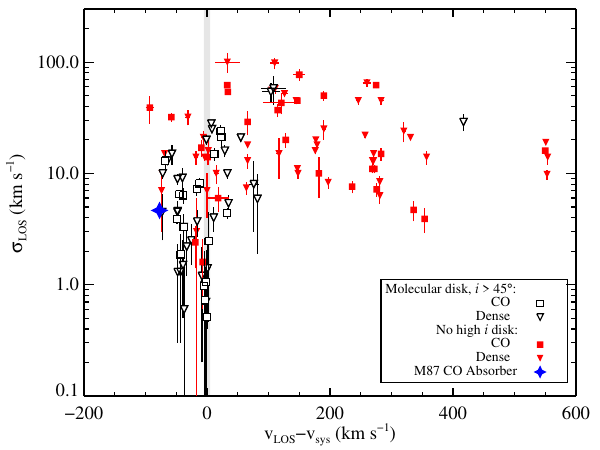}
    \caption{Kinematics for the CO absorber in M87 compared to those of CO or dense-gas tracers against bright nuclei in other massive ETGs \citep[][]{rose20,rose24}. Clouds in other systems are separated into those with more highly inclined molecular disks (open symbols) and those without (filled symbols). In M87, only an atomic gas CND is detected \citep[with $i\sim 25\degr$;][]{osorno23}. The shaded region shows the $cz_\mathrm{obs}$ uncertainty for M87 \citep{cappellari11}.}
    \label{fig:coabs_comp}
\end{figure}

\subsubsection{Kinematics}
\label{sec:disc_abskin}

In a few radio galaxies, CO absorption signatures appear to arise from chance alignment of dense CND molecular clouds and the bright nucleus, especially for cases where molecular CNDs are detected with higher inclination angles \citep[e.g.,][]{boizelle17,ruffa19,rose19b,rose24}. In these cases, absorption-line centroids tend to be close to \vsys\ (within $\pm$50 \kms) as clouds oriented along the disk minor axis have little if any LOS motion. As is shown in Figure~\ref{fig:coabs_comp}, the weighted-average $\overline{v}_\mathrm{LOS} - \vsys = -77.21 \pm 0.44$ \kms\ for the M87 \coone\ and \cotwo transitions are therefore not consistent with clouds in low-ellipticity orbits, although the observed $\overline{\sigma}_\mathrm{LOS} = 4.65\pm0.47$ \kms\ is typical of disk-like clouds. The CO emission-line search provided no convincing support for a molecular phase of the atomic gas disk \citep[oriented at $i\sim 25\degr$;][]{osorno23}, making disk-like obscuration implausible. Compared to clouds not connected with a clear molecular-gas CND, the narrow CO features in M87 reside at the extreme edge of detected features. It may be that such blueshifted clouds with $\tau_\mathrm{CO}\ll 1$ \kms\ are more common generally but only detected for M87 at present due to the uncommon ALMA imaging depth and large number of data sets to reproduce faint absorption.

On kpc scales, cool-core BCGs often host multiple molecular gas clumps and filaments whose gas velocities range between about $\pm$200 \kms\ from the BCG rest frame. Outside of strongly rotationally-broadened regions like the nucleus, CO emission FWHM are typically 30$-$60 \kms. A good example is Abell 1644-S, whose arching CO emission spans $\sim$15 kpc with speeds shifting from about $-$140 to +30 \kms\ relative to its \vsys, with larger-scale flows possibly connecting down to a multi-phase CND. In the most isolated regions, moderate-resolution CO imaging shows narrow emission-line FWHM~$\sim 20$ \kms\ with minimal velocity gradients \citep[][]{Baek2022}, although beam smearing of adjacent clouds likely still appreciably broadens the line intrinsic FWHM \citep[see also][]{Olivares2019,ganguly23}. In NGC 1275, multiple streamers connect CO emission to the CND in a complex fashion \citep{oosterloo24}. In cool-core BCGs, such patchy and filamentary multi-phase gas structures are typical of chaotic cooling of the galactic and circumgalactic medium that greatly increases $\dot{M}_\mathrm{BH}/\dot{M}_\mathrm{Edd}$ in episodic accretion \citep[e.g.,][]{babyk19,gaspari20,pasini21}. The case of Abell 1644-S is also similar to the atomic gas filaments in M87 \citep{osorno23} as the blueshifted clouds connecting to the CND from behind appear to continue in front of the nucleus to be seen in blueshifted absorption \citep{Baek2022,rose24}.

For M87 specifically, \citet{simionescu18} used ALMA imaging to map out CO emission from the dense SE molecular cloud that is embedded within a broader filamentary atomic gas complex $\sim$40\arcsec\ from the nucleus with projected $\vlos - \vsys \sim -129$ \kms. The SE cloud shows a modest gradient of at least a couple $\times$10 \kms\ across the brightest emission-line clump that extends over a projected $\sim$150 pc. At an angular resolution of $\theta_\mathrm{FWHM}\sim 1\farcs4 \times 0\farcs7$ (average $\sim$85 pc), the gas cloud with $S_\mathrm{CO(2-1)} \Delta v = 0.56\pm0.06$ Jy \kms\ and $\vlos-\vsys \sim -129$ \kms\ is resolved into a spherical clump with a diameter of $\sim$100 pc and $\sigma_\mathrm{LOS} \sim 30$ \kms\ together with a smaller, filamentary component with a narrower line width. The latter better matches the observed line widths and low velocity gradient from the CO absorption shown in Figure~\ref{fig:colineabsfit}. Based on the relative distribution of ionized gas and X-ray emission, those authors argue the SE cloud is falling into M87 from behind, with the observed molecular gas surviving passage through the moderately cool-core X-ray gas in this part of the Virgo cluster \citep{shibata01,forman07} only to be destroyed at closer radii either by shocks or by relativistic particles associated with the radio lobes. 

The primary UV/optical absorption lines that are connected to a central outflow are more blueshifted ($\vlos-\vsys \sim -131\pm19$ \kms) and much broader \citep[FWHM~$\lesssim 300$ \kms;][]{tsvetanov99a,sabra03} than is the CO absorption\footnote{Note that these atomic gas velocities were shifted upwards by up to $\sim$60 \kms\ to match Galactic atomic absorption-line centroids (e.g., for \ion{S}{1} or \ion{N}{2}) with the primary \ion{H}{1} centroid at $cz_\mathrm{obs} \sim -10$ \kms\ \citep{davies75}. While higher spectral resolution UV/optical data exist \citep[e.g.,][]{sankrit99}, they do not have sufficient sensitivity to confirm the absorption-line velocity offset from CO and better characterize the line width.}. The velocity centroids of narrow absorption-line gas in AGN winds or outflows may vary by a few \kms\ over several-year timescales \citep[e.g.,][]{misawa14,grier16,misawa19,rose19a,dehghanian24}. If entrained in the same gaseous outflow, the atomic and molecular absorption-line properties for M87 should be more similar \citep{margulis89}. Additionally, two faint optical absorption-line systems were reported at $\vlos \sim 980$ and 1300 \kms\ \citep{carter92,carter97} without any coincident CO absorption.

\subsubsection{Column Densities}
\label{sec:disc_nh2}

The estimated $N_\mathrm{CO} \sim 10^{15}$ cm\pertwo\ is below the detection threshold for many extragalactic ALMA programs, although it is reached by certain Galactic line surveys \citep[e.g.,][]{liszt98}. A standard CO/H$_2$ abundance ratio of $3.2\times 10^{-4}$ \citep{sofia04,bolatto2013} would result in a very low $N_\mathrm{H_2} \sim 1.5\times 10^{18}$ cm\pertwo. That conversion factor is typically applied to systems with peak $N_\mathrm{H_2} \gtrsim 10^{21}$ cm\pertwo\ while such a low $N_\mathrm{CO}$ better comports with more diffuse Galactic lines of sight that have CO/H$_2$ abundances of $\sim$2$\times$10$^{-6}$ \citep{liszt19} to $\sim$8$\times$10$^{-7}$ \citep{pineda10} due to easier CO destruction. These ratios result in our preferred range $N_\mathrm{H_2}/10^{20}\,\mathrm{cm\pertwo} = 0.92^{+0.68}_{-0.36} - 2.3^{+1.7}_{-0.9}$ or $\Sigma_\mathrm{H_2} \sim 0.6-1.4$ \msun\ pc\pertwo\ that matches an estimated total $N_\mathrm{H} \sim (2-5)\times 10^{20}$ cm\pertwo\ towards the M87 nucleus from X-ray spectroscopy \citep[e.g.,][]{wilson02,dimatteo03}.

Simulations suggest CO is created (or persists in an equilibrium state) for gas with $n_\mathrm{H_2} \gtrsim 1000$ cm\perthree\ where dust, H$_2$, and line self-shielding prevents CO dissociation \citep{safranek17}. The low $N_\mathrm{H_2}$ and $A_V \sim 0.04-0.10$ mag estimate \citep{guver09} suggests more limited shielding against interstellar radiation \citep{gnedin14}. Similar to the SE complex, the CO-absorbing gas may be physically located far from the nucleus and be evolving (further) towards atomic gas dominance. The low molecular gas fraction inferred from this $A_V$ range \citep[$f_\mathrm{H_2} \sim 0.1$;][]{snow06} would make \ion{C}{2} and not \ion{C}{1} or CO the dominant reservoir of gas-phase carbon \citep{gerin24}. Far-IR spectroscopy centered about the M87 nucleus does reveal faint [\ion{C}{2}] 158 $\mu$m and [\ion{O}{1}] 63 $\mu$m fine-structure lines \citep[][]{brauher08}, but the $\sim$1.7\arcmin\ aperture diameter (and possible 1\arcmin\ offset from the nucleus) prevents any clear connection to any specific gaseous structure. Adding neutral atomic gas would only give an estimated $N(\mathrm{H}) = 2N(\mathrm{H}_2) \times (1-f_\mathrm{H_2})/f_\mathrm{H_2} \sim (2-4)\times 10^{21}$ cm\pertwo\ $\sim15-30$ \msun\ pc\pertwo.

\begin{figure*}[!th]
    \centering
    \includegraphics[width=\textwidth]{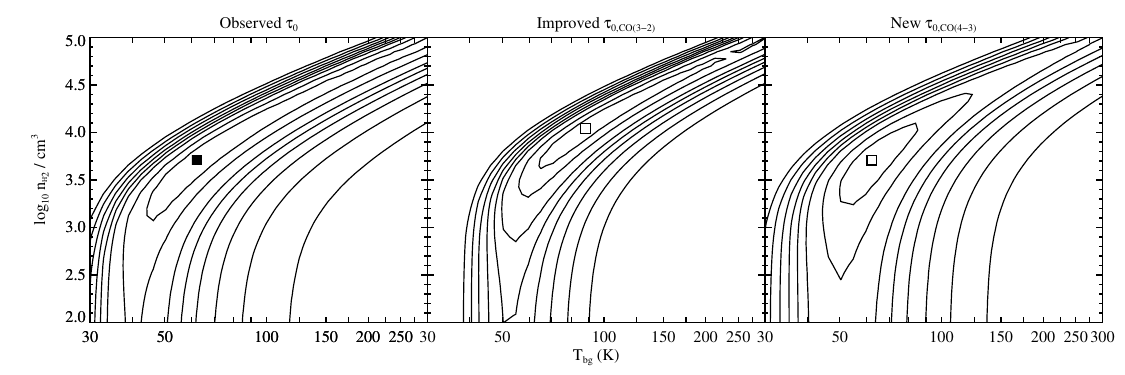}
    \caption{\texttt{radex} radiative transfer modeling using $T_K = 10$ K and a CO line FWHM of 10 \kms. Results are shown for measured (Table~\ref{tbl:narrowabs}; \textit{left panel}) and augmented peak opacities, including an improved $\tau_\mathrm{0,CO(3-2)}$ (\textit{middle}) or a new $\tau_\mathrm{0,CO(4-3)}$ value (\textit{right}). New or improved $\tau_0$ are assumed to reach S/N~$\sim 3.4$, and the $\tau_\mathrm{0,CO(4-3)}$ value is predicted for the best-fitting $n_\mathrm{H_2}$ and $T_\mathrm{bg}$ (\textit{left}; see also Figure~\ref{fig:tex_plot}). Inner $\Delta \chi^2$ contours are at the 1, 2, and 3$\sigma$ level while the remainder are arbitrarily (but linearly) spaced. Either of these new data sets would close the 1$\sigma$ confidence interval (for fixed $T_K$), although \cofour\ observations have the potential to provide the tightest $n_\mathrm{H_2}$ and $T_\mathrm{bg}$ constraints.}
    \label{fig:radex_sim_plot}
\end{figure*}

\subsubsection{Temperatures and Densities}
\label{sec:disc_temp}

The measured $T^{21}_\mathrm{ex} \sim 8\pm2$ K is similar to excitation temperatures found for diffuse Galactic \citep[e.g.,][]{liszt04,roman10} and extragalactic molecular gas. Some massive galaxies have diffuse obscurers with $T_\mathrm{ex} \sim 20-40$ K \citep[][]{israel92,lim17,rose19b,rose19a} while outflow CO line ratios or warmer molecular tracers suggest $T_\mathrm{ex} \sim 30-100$ K and $T_K \sim 400$ K, respectively \citep[][]{dasyra16,cashman21,ruffa22}. Because of the intense interstellar radiation field and cosmic ray environment, the CO absorber is likely to be out of LTE. Indeed, both the measured $\tau_\mathrm{CO(3-2)}$ and inferred $T^{32}_\mathrm{ex} \sim 30\pm 20$ K are elevated over LTE predictions from lower-$J$ CO measurements.

To probe more realistic cloud conditions, we employed the one-dimensional non-LTE radiative transfer code \texttt{radex} \citep{ferland98,van07} to reproduce the observed peak opacities $\overline{\tau}_\mathrm{0,CO(1-0)}$, $\overline{\tau}_\mathrm{0,CO(2-1)}$, and $\tau_\mathrm{0,CO(3-2)}$. \texttt{radex} models excitation of molecular lines in a uniform, isothermal, and homogeneous medium, employing the large velocity gradient approximation to compute photon escape. We assumed the gas is organized in a uniform sphere, although different geometries and escape probabilities give similar results for low-$J$ CO transitions. Radiative transfer modeling typically relies on multiple molecular line tracers and/or isotopologues to probe dense gas conditions. In this case, however, the very low opacity $\tau_\mathrm{0,CO} \ll 1$ simplifies the analysis and allows $^{12}$CO to trace the entire cloud. To be consistent with observations, we adopted $N_\mathrm{CO} = 10^{15}$ cm\pertwo\ from LTE measurements and fixed the CO line FWHM to 10 \kms. The level of gas turbulence is not known, but we assumed a cold gas distribution and tested $T_K = 10 $ K and 30 K in turn. During \texttt{radex} modeling, we explored $n_\mathrm{H_2} = 10^2-10^5$ cm\perthree\ and background temperature $T_\mathrm{bg} = 30-300$ K.

In Figure~\ref{fig:tex_plot}, we compare measured and \texttt{radex} model $\tau_0$ values and present $\chi^2$ contours over a grid of $n_\mathrm{H_2}$ and $T_\mathrm{bg}$ points. These results show strong preference for a very cold distribution ($T_K = 10$ K) with best-fitting $n_\mathrm{H_2} \sim 5100$ cm\perthree\ and $T_\mathrm{bg} \sim 62$ K that is much higher than the $\sim$23 K suggested by the mean interstellar radiation field \citep[ISRF;][]{baes10}, providing additional support for a more nuclear location for the CO absorption. While the 1$\sigma$ contours do not close for reasonable $n_\mathrm{H_2}$ and $T_\mathrm{bg}$, they provide $n_\mathrm{H_2} > 1100$ cm\perthree\ and $T_\mathrm{bg} > 45$ K lower limits. A warmer $T_K = 30$ K gives best-fit $n_\mathrm{H_2} \approx 3.5\times 10^4$ cm\perthree\ and $T_\mathrm{bg} \sim 71$ K with 7$\times$ higher $\chi^2$; still, lower-bound parameter limits are similar to the $T_K = 10$ K case. Based on these \texttt{radex} bounds, the current CO absorption-line data appear more consistent with central $n_\mathrm{H_2} > 1000$ cm\perthree\ for a thin filament ($r_\mathrm{c} < 0.1$ pc) instead of a more diffuse filament or spherical cloud.

Additional higher-frequency ALMA 12-m imaging may break the degeneracy between $n_\mathrm{H_2}$ and $T_\mathrm{bg}$ and help to better constrain $T_K$ \citep[e.g., ][]{vankempen16}. We tested two cases: one that improved the existing $\tau_\mathrm{0,CO(3-2)}$ measurement and the other that obtained new Band 8 imaging to detect $\tau_\mathrm{0,CO(4-3)}$. In both cases, we assumed that the new peak opacity reached the same fractional precision as the Xa1 $\tau_\mathrm{0,CO(1-0)}$ measurement (S/N~$\sim 3.4$). In Figure~\ref{fig:radex_sim_plot}, these \texttt{radex} results show closed 1$\sigma$ confidence intervals in both cases, although adding in the \cofour\ measurement to the values in Table~\ref{tbl:narrowabs} provides much tighter constraints. In the best case, $n_\mathrm{H_2}$ and $T_\mathrm{bg}$ could be known to within $\sim$0.3 dex and 10$-$20 K, respectively. Using the \texttt{OT}, we estimate these new $\tau_0$ measurements would require a minimum on-source (total) time of 56 min (112 min) for the Band 7 imaging while for Band 8 the require time would be up to 6.25 hrs (12.1 hrs).

\subsubsection{Origin of the CO Absorption}

Now, we employ the measured CO line kinematics, the estimated column densities, and non-LTE modeling results to explore both gravitationally bound, spherical cloud and filament origins for the CO absorption features. Based on an isolated absorption line and narrow widths, we discount any interpretation of spatially diffuse molecular gas. We appeal to both Galactic and extragalactic observations for comparison but do not find a fully coherent explanation for the measured $\sigmalos$ and inferred $T_K$, $N_\mathrm{H_2}$ and $n_\mathrm{H_2}$ values.

A cloud is often considered virialized when parameter $\alpha_\mathrm{vir} \equiv 5\sigma^2/\pi GR\Sigma \approx 1$, although real clouds in luminous galaxy centers are centered on $\alpha_\mathrm{vir} \approx 1.7$ \citep{choi24}. Correlations between cloud size $R$ and emission-line width for gravitationally-bound clouds give a range of justifiable results. From the fitted absorption-line $\overline{\sigma}_\mathrm{LOS}$, Galactic or LMC studies \citep[Larson's relations; e.g.,][]{solomon87,wong22} are consistent with a cloud radius $r_\mathrm{cloud} \sim 20-40$ pc while rescaled relations in large-scale CNDs in luminous ETGs suggest $r_\mathrm{cloud} \sim 10-20$ pc \citep[e.g.,][]{utomo15,rosolowsky21}. From turbulent clouds in the Galactic center (using dense-gas tracers), the observed \sigmalos\ cloud be matched to an $r_\mathrm{cloud} \sim 2-4$ pc \citep{shetty12}. We note that absorption probes only a thin line through the cloud, so a CO emission-line width would likely be broader. Assuming $\alpha_\mathrm{vir} \approx 1.7$, the $r_\mathrm{cloud} \sim 4-30$ pc range and the measured $\overline{\sigma}_\mathrm{LOS}$ would require an implausibly high gas surface mass density $\Sigma_\mathrm{H_2} \sim 400-7000$ \msun\ pc\pertwo\ considering the CO emission-line non-detection discussed in Section~\ref{sec:disc_em} and the absorption-line $\Sigma_\mathrm{H_2} \sim 1$ \msun\ pc\pertwo\ estimate. If the absorbing cloud is well centered on the AGN and uniformly distributed, the $N_\mathrm{H_2}$ would correspond to $\overline{n}_\mathrm{H_2} \sim 0.5-9$ cm\perthree\ with a total $M_\mathrm{H_2} \sim 50-7000$ \msun. For comparison, the more compact region of the SE cloud has an estimated $\overline{n}_\mathrm{H_2} \sim 50$ cm\perthree. Adopting typical cloud radial density profiles and small core sizes \citep[$\sim$0.1 pc; e.g.,][]{csengeri17,andre22} relative to the $r_\mathrm{cloud}$ range above, the expected \textit{central} densities of 100$-$1800 cm\perthree\ would be lower than typical dense gas cores as well as the \coone\ critical density \citep[since the opacity $\tau \ll 1$;][]{scoville12}.

More likely, this CO absorption originates in a foreground filament. Simulations suggest that accreting clouds like those composing the SE complex or the dense dusty clump are preferentially drawn into narrow filaments by magnetic fields that thread massive galaxies \citep[e.g.,][]{fournier24}. The $N_\mathrm{H_2}$ range is much lower than typical Galactic filament peak values \citep{arzoumanian11,arzoumanian18,konyves20} but better matches properties of extragalactic filamentary structures \citep[e.g.,][]{sancisi08,blok14}. In Galactic systems, resolved filamentary structures seen in emission have typical line widths $\lesssim$1 \kms\ and narrow core widths $r_\mathrm{c} \lesssim 0.1$ pc \citep[wherein $n_\mathrm{H_2}$ is uniform; e.g.,][]{arzoumanian11,koch15,panopoulou16,rivera16,indebetouw20}. For the measured $\overline{\sigma}_\mathrm{LOS}$, filament size-emission-line width relations would be consistent with an effective $r_\mathrm{c} \sim 0.5$ pc or more, which is at the very upper limit for Galactic filaments. However, the MW filament environment is very different than for the M87 nucleus, and the latter has ionized gas filament intrinsic widths of $\sim$200$-$300 pc \citep[][]{sparks93,sparks09,forman07}. We note that within and around radio galaxies, even more diffuse filaments reaching $\sim$kpc widths have been observed \citep[][]{yusef22}.

Calculating $\overline{n}_\mathrm{H_2}$ through the center of a Plummer filament model with matching $N_\mathrm{H_2}$ \citep{rivera16}, the peak H$_2$ densities should reach \textit{at least} 1000$-$2500 cm\perthree, which is consistent with simulations of filament formation \citep{hennebelle13}. If $r_\mathrm{c} \sim 0.1$ pc, the corresponding linear mass $M_\mathrm{H_2}/\ell \sim 0.5-1$ \msun\ pc\per\ would be at the very bottom of the linear mass distribution for Galactic systems associated with star formation \citep[e.g.,][]{sancisi08}, although it does match more diffuse extragalactic filaments associated with accretion \citep[e.g.,][]{blok14}. In any case, the M87 absorber would be in a very distinct region of the \sigmalos\ vs.\ $M_\mathrm{H_2}/\ell$ or $N_\mathrm{H_2}$ parameter space. The observed $\overline{\sigma}_\mathrm{LOS}$ is an order-of-magnitude higher than line widths for Galactic systems while having two orders of magnitude lower $N_\mathrm{H_2}$ \cite[based on dense-gas tracers;][]{arzoumanian13,chen24}. If positioned slightly off-center from the M87 nucleus, such a filament would allow for higher peak $n_\mathrm{H_2}$ and $M_\mathrm{H_2}/\ell$ values.

The estimated $\Sigma_\mathrm{H_2} \sim 1$ \msun\ pc\pertwo\ is far below the threshold needed to reconcile the measured $\overline{\sigma}_\mathrm{LOS}$ with a virialized, compact filament. Instead, interactions with the hot interstellar medium or an envelope of warm, partially ionized gas may provide sufficient pressure to confine the CO absorber. In the first case for an isothermal X-ray-emitting distribution, the upper-bound electron density $n_e \sim 0.5$ cm\perthree\ in the region of large-scale filaments \citep{olivares25} and temperature $kT_e \sim 1.5-2$ keV \citep[$T_e/10^6 \,\mathrm{K} \sim 15-25$;][]{sparks04,forman07} give rise to a pressure $P \gtrsim (1-2)\times 10^{-9}$ dynes cm\pertwo. Pressure from the X-ray emission of filament regions is a only factor of 2$-$3 lower \citep{olivares25}. In the second case, warm ionized gas with $T_e \sim 10^4$ K could provide similar pressure if $n_e \sim 1600$ cm\perthree\ (\citealp{ford79}; see also \citealp{heckman89}). For comparison, the cloud or filament scenarios with their corresponding $\overline{n}_\mathrm{H_2}$ and a $T_K$ reaching as high as $\sim$25$-$50 K results in internal $P \ll 10^{-10}$ dynes cm\pertwo, leading to pressure confinement of the diffuse molecular gas. Without more extensive line measurements and full radiative transfer modeling, it is not clear how a thin filament embedded in such an extreme environment could retain a low $T_K \sim 10$ K.

\section{Conclusion}
\label{sec:conc}

We conducted a CO line search of M87 using archival ALMA 12-m imaging, focusing on the CND and the dusty clouds and filaments seen in optical imaging in the inner $\sim$kpc. Despite the deep ALMA imaging and multiple spws containing redshifted CO transitions, we did not clearly identify any CO emission. Over the atomic gas CND, the upper limit $M_\mathrm{H_2} < 2.3\times 10^5$ \msun\ is $\sim$20$\times$ lower than achieved in previous CO surveys. In other regions containing atomic gas filaments and dust, we placed upper limits of $M_\mathrm{H_2} \lesssim 10^5 - 10^6$ \msun\ with a global $<$4$\times$10$^6$ \msun\ limit for a circular aperture with $R<2$ kpc. In narrow gaseous regions, potential CO emission-line signatures in extracted line profiles do not arise from compact sources. 

During this search, we confidently detected molecular gas in absorption against the bright nucleus in the \coone, \cotwo, and \cothree\ transitions. The absorber has very low opacity with average peak $\overline{\tau}_\mathrm{0,CO(1-0)} = 0.00239 \pm 0.00054$ and integrated $\overline{\tau}_\mathrm{CO(1-0)} = 0.0281 \pm 0.0045$ \kms. Measured opacities are slightly higher for the higher-$J$ absorption lines. Line fitting showed a consistently narrow, slightly blueshifted feature with average $\overline{v}_\mathrm{LOS} - \vsys \sim -77.2\pm 0.4$ \kms\ and $\overline{\sigma}_\mathrm{LOS} \sim 4.7\pm 0.5$ \kms. From multiple low-S/N CO absorption lines we calculated column densities $N_\mathrm{CO} = \left(1.16^{+0.21}_{-0.28}\right) \times 10^{15}$ cm\pertwo\ and $N_\mathrm{H_2}/10^{20}\,\mathrm{cm}\pertwo = 0.92^{+0.68}_{-0.36} - 2.3^{+1.7}_{-0.9}$ (1$\sigma$ uncertainties) assuming the gas is in a diffuse molecular phase. These molecular absorption features appear inconsistent with blueshifted atomic gas absorption detected in UV/optical spectroscopy at a relative speed of about $-$130 \kms\ with line FWHM~$\lesssim 300$ \kms, and we disfavor interpreting this CO absorber as arising from molecular gas entrained in an outflow.

Instead, the CO absorption-line properties are more consistent with a filament in a chance alignment with the AGN core, like those seen close in projection to the nucleus. Based on opacity measurements, LTE relations suggest the molecular gas is cold with excitation temperatures $T^{21}_\mathrm{ex} = 7.77^{+1.34}_{-1.94}$ K (1$\sigma$) and $T^{32}_\mathrm{ex} \sim 30$ K, hinting at a $J=2$ population that is elevated over LTE predictions. Non-LTE radiative transfer modeling finds much better agreement with the observed $\tau_0$ values, and \texttt{radex} prefers a low kinematic temperature $T_K \sim 10$ K and more dense $n_\mathrm{H_2} \sim 5000$ cm\perthree\ that is not consistent with a spherical, gravitationally bound cloud for the measured $\overline{\sigma}_\mathrm{LOS}$. The surrounding background radiation field with best-fit $T_\mathrm{bg} \sim 60$ K is hotter than expected from the mean ISRF, suggesting the CO absorber lies close to the AGN. This (uniform) $n_\mathrm{H_2}$ estimate better matches expectations for a thin, pressure-confined filament, perhaps viewed slightly offset from the compact core location.

\begin{acknowledgments}
    We thank the anonymous reviewer for helpful suggestions. We also thank Aaron Barth, Andrew Baker, Jeremy Darling, and Luis Ho for advice and encouragement. We are grateful for advice from Ivan Marti-Vidal, Ciriaco Goddi, and Hiroshi Nagai, together with the ALMA NAASC, in dealing with APP-mode ALMA data sets and bandpass calibration limitations.

    Partial support for this work was provided by the NSF through award SOSPADA-003 from the NRAO. XL, NL, BJD, and JD thank the Brigham Young University College of Computational, Mathematical, and Physical Sciences and the Department of Physics and Astronomy for additional research funding.

    This research is based, in part, on observations made with the NASA/ESA Hubble Space Telescope obtained from the Space Telescope Science Institute, which is operated by the Association of Universities for Research in Astronomy, Inc., under NASA contract NAS 5-26555. These observations are associated with programs GO-12271, GO-13731, and GO-14256. This paper makes use of the following ALMA data in ADS/JAO.ALMA\# codes: 2011.0.00001.CAL, 2012.1.00661.S, 2013.1.00073.S, 2013.1.01022.S, 2015.1.00030.S, 2015.1.01170.S, 2015.1.01352.S, 2016.1.00021.S, 2016.1.01154.V, 2017.1.00608.S, 2017.1.00841.V, 2017.1.00842.V, 2019.1.00807.S, and 2021.1.01398.S. ALMA is a partnership of ESO (representing its member states), NSF (USA) and NINS (Japan), together with NRC (Canada), NSTC and ASIAA (Taiwan), and KASI (Republic of Korea), in cooperation with the Republic of Chile. The Joint ALMA Observatory is operated by ESO, AUI/NRAO and NAOJ. The National Radio Astronomy Observatory is a facility of the National Science Foundation operated under cooperative agreement by Associated Universities, Inc.
\end{acknowledgments}

\facilities{HST (ACS,WFC3), ALMA, ACA}

\software{\texttt{CASA} \citep{mcmullin07}, \texttt{GALFIT} \citep{peng02}, \texttt{radex} \citep{van07}, ALMA \texttt{OT} \citep{immer24}, \texttt{PlotDigitizer} \citep{plotdig}}

\appendix
\restartappendixnumbering
\renewcommand{\thetable}{A\arabic{table}}

\begin{deluxetable}{ccrrcccccccc}[!ht]
\tabletypesize{\scriptsize}
\tablecaption{M87 ALMA 12-m Observations and Imaging Properties\label{tbl:sample}}
\tablewidth{0pt}
\tablehead{
\colhead{Obs.} & \colhead{Member} & \colhead{$\nu_\mathrm{obs}$} & \colhead{$\Delta\nu_\mathrm{obs}$} & \colhead{Baselines} & \colhead{$\theta_\mathrm{maj}\times\theta_\mathrm{min}$} & \colhead{MRS} & \colhead{Time} & \colhead{Peak $S_\nu$} & \colhead{MFS rms} & \colhead{PWV} & \colhead{Pol.} \\[-1.5ex]
\colhead{Date} & \colhead{OUS} & \colhead{(GHz)} & \colhead{(MHz)} & \colhead{(km)} & \colhead{$\theta_\mathrm{PA}$} & \colhead{(arcsec)} & \colhead{(hrs)} & \colhead{(Jy)} & \colhead{(mJy beam$^{-1}$)} & \colhead{(mm)} & \\[-1.5ex]
\colhead{(1)} & \colhead{(2)} & \colhead{(3)} & \colhead{(4)} & \colhead{(5)} & \colhead{(6)} & \colhead{(7)} & \colhead{(8)} & \colhead{(9)} & \colhead{(10)} & \colhead{(11)} & \colhead{(12)}
}
\startdata
    \multicolumn{12}{c}{\textbf{2012.1.00661.S} (PI: Vlahakis)} \\ \cline{1-12}
     &  & 99.90$-$101.89 & 15.63 & \multirow{4}{*}{0.015$-$0.423} &  & \multirow{4}{*}{12.5} & \multirow{4}{*}{1.71} & \multirow{4}{*}{2.10} & \multirow{4}{*}{1.81} & \multirow{4}{*}{1.41} & \multirow{4}{*}{XX YY} \\
    2014 & A002/X6444ba & 101.78$-$103.76 & 15.63 &  & $2\farcs03\times1\farcs42$ &  &  &  &  &  &  \\
    7 Mar & /X1b0 & 111.92$-$113.91 & 15.63 &  & $60.2^\circ$ &  &  &  &  &  &  \\
     &  & 113.85$-$115.72 & 0.49 &  &  &  &  &  &  &  &  \\[1.5ex]
     &  & 331.50$-$333.48 & 15.63 & \multirow{4}{*}{0.015$-$0.347} &  & \multirow{4}{*}{6.04} & \multirow{4}{*}{1.89} & \multirow{4}{*}{1.01} & \multirow{4}{*}{0.104} & \multirow{4}{*}{0.97} & \multirow{4}{*}{XX YY} \\
    2014 & A002/X6444ba & 333.30$-$335.29 & 15.63 &  & $1\farcs16\times0\farcs60$ &  &  &  &  &  &  \\
    28 Jan & /X1b4 & 343.41$-$345.28 & 9.77 &  & $-$83.7$^\circ$ &  &  &  &  &  &  \\
     &  & 345.30$-$347.29 & 15.63 &  &  &  &  &  &  &  &  \\[0.5ex]  \cline{1-12}
    \multicolumn{12}{c}{\textbf{2013.1.00073.S} (PI: Tan)} \\ \cline{1-12}
     &  & 213.00$-$214.99 & 15.63 & \multirow{4}{*}{0.021$-$0.784} &  & \multirow{4}{*}{4.04} & \multirow{4}{*}{0.25} & \multirow{4}{*}{1.75} & \multirow{4}{*}{1.10} & \multirow{4}{*}{0.69} & \multirow{4}{*}{XX YY} \\
    2015 & A001/X12f & 215.00$-$216.99 & 15.63 &  & $0\farcs54\times0\farcs44$ &  &  &  &  &  &  \\
    14 Jun & /X20f & 228.63$-$230.50 & 0.49 &  & $-$47.8$^\circ$ &  &  &  &  &  &  \\
     &  & 230.57$-$232.56 & 15.63 &  &  &  &  &  &  &  &  \\[1.5ex]
     &  & 213.00$-$214.99 & 15.63 & \multirow{4}{*}{0.043$-$1.574} &  & \multirow{4}{*}{1.88} & \multirow{4}{*}{0.50} & \multirow{4}{*}{0.93} & \multirow{4}{*}{1.99} & \multirow{4}{*}{0.74} & \multirow{4}{*}{XX YY} \\
    2015 & A001/X12f & 215.00$-$216.99 & 15.63 &  & $0\farcs27\times0\farcs22$ &  &  &  &  &  &  \\
    16 Aug & /X20d & 228.63$-$230.50 & 0.49 &  & 14.3$^\circ$ &  &  &  &  &  &  \\
     &  & 230.57$-$232.56 & 15.63 &  &  &  &  &  &  &  &  \\[0.5ex]  \cline{1-12}
    \multicolumn{12}{c}{\textbf{2013.1.01022.S} (PI: Asada)} \\ \cline{1-12}
     &  & 89.51$-$91.48 & 31.25 & \multirow{4}{*}{0.041$-$2.070} &  & \multirow{4}{*}{23.8} & \multirow{4}{*}{3.32} & \multirow{4}{*}{2.47} & \multirow{4}{*}{0.27} & \multirow{4}{*}{1.56} &  \\
    2015 & A001/X13a & 91.45$-$93.42 & 31.25 &  & $0\farcs46\times0\farcs38$ &  &  &  &  &  & XX XY \\
    19 Sep & /Xbe & 101.51$-$103.48 & 31.25 &  & 30.0$^\circ$ &  &  &  &  &  & YX YY \\
     &  & 103.51$-$105.48 & 31.25 &  &  &  &  &  &  &  &  \\[0.5ex]  \cline{1-12}
    \multicolumn{12}{c}{\textbf{2015.1.00030.S} (PI: Vlahakis)} \\ \cline{1-12}
     &  & 99.88$-$101.86 & 15.63 & \multirow{4}{*}{0.085$-$16.20} &  & \multirow{4}{*}{0.97} & \multirow{4}{*}{0.66} & \multirow{4}{*}{1.96} & \multirow{4}{*}{0.51} & \multirow{4}{*}{0.96} & \multirow{4}{*}{XX YY} \\
    2015 & A001/X2d8 & 101.87$-$103.86 & 15.63 &  & $0\farcs10\times0\farcs046$ &  &  &  &  &  &  \\
    8 Nov & /X5 & 112.28$-$114.16 & 0.98 &  & $-$56.5$^\circ$ &  &  &  &  &  &  \\
     &  & 113.85$-$115.72 & 0.98 &  &  &  &  &  &  &  &  \\[0.5ex] \cline{1-12}
    \multicolumn{12}{c}{\textbf{2015.1.01170.S} (PI: Asada)} \\ \cline{1-12}
     &  & 89.50$-$91.47 & 31.25 & \multirow{4}{*}{0.085$-$16.20} &  & \multirow{4}{*}{0.88} & \multirow{4}{*}{2.29} & \multirow{4}{*}{1.94} & \multirow{4}{*}{0.34} & \multirow{4}{*}{1.34} &  \\
    2015 & A001/X2df & 91.44$-$93.41 & 31.25 &  & $0\farcs07\times0\farcs05$ &  &  &  &  &  & XX XY \\
    11 Nov & /X135 & 101.50$-$103.47 & 31.25 &  & 34.8$^\circ$ &  &  &  &  &  & YX YY \\
     &  & 103.50$-$105.47 & 31.25 &  &  &  &  &  &  &  &  \\[0.5ex] \cline{1-12}
    \multicolumn{12}{c}{\textbf{2015.1.01352.S} (PI: Doi)} \\ \cline{1-12}
     &  & 89.50$-$91.48 & 15.63 & \multirow{4}{*}{0.253$-$16.20} &  & \multirow{4}{*}{0.90} & \multirow{4}{*}{0.034} & \multirow{4}{*}{1.75} & \multirow{4}{*}{0.69} & \multirow{4}{*}{0.97} & \multirow{4}{*}{XX YY} \\
    2015 & A001/X2d6 & 91.43$-$93.42 & 15.63 &  & $0\farcs06\times0\farcs05$ &  &  &  &  &  &  \\
    27 Oct & /X2ba & 101.50$-$103.48 & 15.63 &  & 39.3$^\circ$ &  &  &  &  &  &  \\
     &  & 103.50$-$105.48 & 15.63 &  &  &  &  &  &  &  &  \\[1.5ex]
     &  & 136.99$-$138.98 & 15.63 & \multirow{4}{*}{0.068$-$15.24} &  & \multirow{4}{*}{0.72} & \multirow{4}{*}{0.36} & \multirow{4}{*}{1.72} & \multirow{4}{*}{0.35} & \multirow{4}{*}{1.81} & \multirow{4}{*}{XX YY} \\
    2015 & A001/X2d6 & 138.93$-$140.91 & 15.63 &  & $0\farcs06\times0\farcs05$ &  &  &  &  &  &  \\
    31 Oct & /X2be & 148.99$-$150.98 & 15.63 &  & $-$37.2$^\circ$ &  &  &  &  &  &  \\
     &  & 150.99$-$152.98 & 15.63 &  &  &  &  &  &  &  &  \\[1.5ex]
     &  & 222.99$-$224.98 & 15.63 & \multirow{4}{*}{0.068$-$15.24} &  & \multirow{4}{*}{1.34} & \multirow{4}{*}{0.22} & \multirow{4}{*}{1.25} & \multirow{4}{*}{0.47} & \multirow{4}{*}{1.24} & \multirow{4}{*}{XX YY} \\
    2015 & A001/X2d6 & 224.99$-$226.98 & 15.63 &  & $0\farcs03\times0\farcs03$ &  &  &  &  &  &  \\
    31 Oct & /X2c2 & 238.98$-$240.97 & 15.63 &  & $-$35.1$^\circ$ &  &  &  &  &  &  \\
     &  & 240.98$-$242.97 & 15.63 &  &  &  &  &  &  &  &  \\[0.5ex] \cline{1-12}
\enddata
\begin{singlespace}
  \tablecomments{\footnotesize Properties of ALMA 12-m interferometric and phased-array (APP-mode; .V) data sets that were centered on the M87 nucleus, with the project code and Principle Investigator listed above each synopsis. Cols.\ (1) and (2) give the UT observation date for each member OUS, respectively, while cols.\ (3) and (4) give the frequency coverage of each spectral window (spw) along with the associated channel binning. For FDM spws, online Hanning smoothing decreases the effective spectral resolution by a factor of up to two. Unprojected baseline ranges in col.\ (5) give rise to the corresponding synthesized beam parameters from the final MFS image in col.\ (6) and the maximum recoverable scale (MRS; adopted from the ALMA Science Archive) in col.\ (7). Col.\ (8) reports the on-source integration times, including latencies. Cols.\ (9) and (10) give the peak continuum flux density and rms of the MFS image at the average frequency for the four spws after applying self-calibration loops, if possible. Col.\ (11) gives the average precipitable water vapor (PWV; adopted from the ALMA Science Archive) over that time interval. Col.\ (12) reports the dual (XX YY) or cross-polarization (XY YX) setup.}
\end{singlespace}
\end{deluxetable}

\setcounter{table}{0}
\begin{deluxetable}{ccrrcccccccc}[!ht]
\tabletypesize{\scriptsize}
\tablecaption{M87 ALMA 12-m Observations and Imaging Properties}
\tablewidth{0pt}
\tablehead{
\colhead{Obs.} & \colhead{Member} & \colhead{$\nu_\mathrm{obs}$} & \colhead{$\Delta\nu_\mathrm{obs}$} & \colhead{Baselines} & \colhead{$\theta_\mathrm{maj}\times\theta_\mathrm{min}$} & \colhead{MRS} & \colhead{Time} & \colhead{Peak $S_\nu$} & \colhead{MFS rms} & \colhead{PWV} & \colhead{Pol.} \\[-1.5ex]
\colhead{Date} & \colhead{OUS} & \colhead{(GHz)} & \colhead{(MHz)} & \colhead{(km)} & \colhead{$\theta_\mathrm{PA}$} & \colhead{(arcsec)} & \colhead{(hrs)} & \colhead{(Jy)} & \colhead{(mJy beam$^{-1}$)} & \colhead{(mm)} & \\[-1.5ex]
\colhead{(1)} & \colhead{(2)} & \colhead{(3)} & \colhead{(4)} & \colhead{(5)} & \colhead{(6)} & \colhead{(7)} & \colhead{(8)} & \colhead{(9)} & \colhead{(10)} & \colhead{(11)} & \colhead{(12)}
}
\startdata
    \multicolumn{12}{c}{\textbf{2016.1.00021.S} (PI: Vlahakis)} \\ \cline{1-12}
     &  & 112.00$-$113.99 & 15.63 & \multirow{4}{*}{0.021$-$3.64} &  & \multirow{4}{*}{3.16} & \multirow{4}{*}{1.59} & \multirow{4}{*}{1.73} & \multirow{4}{*}{0.37} & \multirow{4}{*}{0.57} & \multirow{4}{*}{XX YY} \\
    2017 & A001/X899 & 112.62$-$114.49 & 0.98 &  & $0\farcs22\times0\farcs21$ &  &  &  &  &  &  \\
    15 Aug & /Xa1 & 113.85$-$115.72 & 0.98 &  & $-$64.8$^\circ$ &  &  &  &  &  &  \\
     &  & 114.04$-$115.91 & 0.98 &  &  &  &  &  &  &  &  \\[0.5ex] \cline{1-12}
    \multicolumn{12}{c}{\textbf{2016.1.01154.V} (PI: EHT Consortium)} \\ \cline{1-12}
     &  & 212.17$-$214.04 & 7.81 & \multirow{4}{*}{0.015$-$0.375} &  & \multirow{4}{*}{12.87} & \multirow{4}{*}{1.22} & \multirow{4}{*}{1.18} & \multirow{4}{*}{1.02} & \multirow{4}{*}{0.81} &  \\
    2017 & A001/X11a7 & 214.17$-$216.04 & 7.81 &  & $2\farcs10\times0\farcs98$ &  &  &  &  &  & XX XY \\
    5 Apr & /X3a & 226.17$-$228.04 & 7.81 &  & $-$82.0$^\circ$ &  &  &  &  &  & YX YY \\
     &  & 228.17$-$230.04 & 7.81 &  &  &  &  &  &  &  &  \\[1.5ex]
     &  & 212.17$-$214.04 & 7.81 & \multirow{4}{*}{0.015$-$0.279} &  & \multirow{4}{*}{9.95} & \multirow{4}{*}{0.89} & \multirow{4}{*}{1.14} & \multirow{4}{*}{0.54} & \multirow{4}{*}{0.69} &  \\
    2017 & A001/X11a7 & 214.17$-$216.04 & 7.81 &  & $2\farcs17\times1\farcs50$ &  &  &  &  &  & XX XY \\
    6 Apr & /X3c & 226.17$-$228.04 & 7.81 &  & $-$76.0$^\circ$ &  &  &  &  &  & YX YY \\
     &  & 228.17$-$230.04 & 7.81 &  &  &  &  &  &  &  &  \\[1.5ex]
     &  & 212.17$-$214.04 & 7.81 & \multirow{4}{*}{0.015$-$0.161} &  & \multirow{4}{*}{12.90} & \multirow{4}{*}{1.69} & \multirow{4}{*}{1.32} & \multirow{4}{*}{1.14} & \multirow{4}{*}{1.11} &  \\
    2017 & A002/Xbee37d & 214.17$-$216.04 & 7.81 &  & $1\farcs18\times0\farcs80$ &  &  &  &  &  & XX XY \\
    11 Apr & /Xb & 226.17$-$228.04 & 7.81 &  & 70.5$^\circ$ &  &  &  &  &  & YX YY \\
     &  & 228.17$-$230.04 & 7.81 &  &  &  &  &  &  &  &  \\[0.5ex] \cline{1-12}
    \multicolumn{12}{c}{\textbf{2017.1.00608.S} (PI: Marti-Vidal)} \\ \cline{1-12}
     &  & 211.51$-$213.48 & 31.25 & \multirow{4}{*}{0.015$-$1.398} &  & \multirow{4}{*}{3.70} & \multirow{4}{*}{0.58} & \multirow{4}{*}{0.93} & \multirow{4}{*}{1.68} & \multirow{4}{*}{0.77} &  \\
    2018 & A001/X129e & 213.51$-$215.48 & 31.25 &  & $0\farcs41\times0\farcs26$ &  &  &  &  &  & XX XY \\
    25 Sep & /X1d & 227.51$-$229.48 & 31.25 &  & $-$53.1$^\circ$ &  &  &  &  &  & YX YY \\
     &  & 229.51$-$231.48 & 31.25 &  &  &  &  &  &  &  &  \\[0.5ex] \cline{1-12}
    \multicolumn{12}{c}{\textbf{2017.1.00841.V} (PI: Doeleman)} \\ \cline{1-12}
     &  & 212.17$-$214.04 & 7.81 & \multirow{4}{*}{0.015$-$0.418} &  & \multirow{4}{*}{7.66} & \multirow{4}{*}{1.78} & \multirow{4}{*}{0.709} & \multirow{4}{*}{2.95} & \multirow{4}{*}{2.06} &  \\
    2018 & A001/X12d1 & 214.17$-$216.04 & 7.81 &  & $0\farcs89\times0\farcs79$ &  &  &  &  &  & XX XY \\
    20 Apr & /X25 & 226.17$-$228.04 & 7.81 &  & 46.1$^\circ$ &  &  &  &  &  & YX YY \\
     &  & 228.17$-$230.04 & 7.81 &  &  &  &  &  &  &  &  \\[1.5ex]
     &  & 212.17$-$214.04 & 7.81 & \multirow{4}{*}{0.015$-$0.418} &  & \multirow{4}{*}{7.66} & \multirow{4}{*}{1.29} & \multirow{4}{*}{0.72} & \multirow{4}{*}{3.15} & \multirow{4}{*}{2.10} &  \\
    2018 & A001/X12d1 & 214.17$-$216.04 & 7.81 &  & $0\farcs95\times0\farcs75$ &  &  &  &  &  & XX XY \\
    21 Apr & /X27 & 226.17$-$228.04 & 7.81 &  & 40.6$^\circ$ &  &  &  &  &  & YX YY \\
     &  & 228.17$-$230.04 & 7.81 &  &  &  &  &  &  &  &  \\[1.5ex]
     &  & 212.17$-$214.04 & 7.81 & \multirow{4}{*}{0.015$-$0.418} &  & \multirow{4}{*}{7.66} & \multirow{4}{*}{1.68} & \multirow{4}{*}{0.84} & \multirow{4}{*}{2.27} & \multirow{4}{*}{2.67} &  \\
    2018 & A001/X12d1 & 214.17$-$216.04 & 7.81 &  & $0\farcs88\times0\farcs82$ &  &  &  &  &  & XX XY \\
    24 Apr & /X255 & 226.17$-$228.04 & 7.81 &  & 11.0$^\circ$ &  &  &  &  &  & YX YY \\
     &  & 228.17$-$230.04 & 7.81 &  &  &  &  &  &  &  &  \\[0.5ex] \cline{1-12}
    \multicolumn{12}{c}{\textbf{2017.1.00842.V} (PI: Lu)} \\ \cline{1-12}
     &  & 85.34$-$87.20 & 7.81 & \multirow{4}{*}{0.015$-$0.372} &  & \multirow{4}{*}{19.39} & \multirow{4}{*}{1.51} & \multirow{4}{*}{1.38} & \multirow{4}{*}{0.73} & \multirow{4}{*}{2.88} &  \\
    2018 & A001/X12d0 & 87.34$-$89.20 & 7.81 &  & $2\farcs57\times1\farcs99$ &  &  &  &  &  & XX XY \\
    14 Apr & /Xe & 97.40$-$99.26 & 7.81 &  & 4.3$^\circ$ &  &  &  &  &  & YX YY \\
     &  & 99.34$-$101.20 & 7.81 &  &  &  &  &  &  &  &  \\[0.5ex]
\enddata
\begin{singlespace}
  \tablecomments{\footnotesize cont.}
\end{singlespace}
\end{deluxetable}

\section{Continuum Properties}
\label{app:cont}

We summarize ALMA 12-m sample properties in Table~\ref{tbl:sample}, including the peak continuum intensity $S_\nu$ and MFS rms for each member OUS. In this section, we provide a brief description of the continuum-only data products and an upper limit to the dust thermal contributions at mm/sub-mm wavelengths.

\subsection{MFS imaging}
\label{app:mfs}

Representative ALMA 12-m MFS imaging in Figure~\ref{fig:mfs} captures the approaching jet and the double radio lobes in compact configurations while the extended configuration data resolve the central continuum from the collimated jet that is prominent in both lower-frequency VLBI and HST imaging \citep[e.g., see][]{reid89,biretta99,walker18}. High-resolution Band 6 imaging isolates features M (sometimes referred to as L) and N1 that lie closer to the nucleus while N2 seen in VLBI data is only detected by ALMA as a slight elongation from the core when using uniform weighting. Notably, these data do not reveal (or are not sufficiently sensitive to detect) diffuse non-thermal emission beyond M that leads into the superluminal HST-1 clump at a distance of $\sim$1\arcsec\ \citep[at least in standard robust $r=0.5$ weighting; see also][]{giroletti12,park19}. From intermediate-resolution MFS imaging, we identify clumps A--F and I in the approaching jet.

We fit the X2c2 MFS image (with uniform weighting; $\theta_\mathrm{maj} \times \theta_\mathrm{min} = 27 \times 21$ mas$^2$ or a FWHM~$\sim 2$ pc) using the \texttt{CASA} \texttt{imfit} task with a nuclear source and an additional, offset elliptical Gaussian to characterize the N2 clump. In the image-plane fit, the dominant source distribution is identical to the synthesized beam while the N2 component has observed FWHM axes of 49.5$\times$18.5 mas$^2$, appearing to be resolved only in the jet direction along a $\mathrm{PA} = -75.6\degr$. While heavily affected by beam smearing, the N2 component centroid is offset from the nucleus by ($\Delta$R.A.,$\Delta$decl.) = ($-$20.0,+5.3) mas, which is consistent within the formal \texttt{imfit} uncertainties to its location in 22 GHz imaging taken just a few months later \citep{park19}. The \texttt{imfit} components fitted to the core and the N2 clump have \textit{integrated} flux densities of 1.220 Jy and 50.3 mJy, respectively, at 232.98 GHz. \texttt{imfit} residuals do not support any additional sources within the inner 50 mas with a background rms of $\sim$0.83 mJy beam\per. The residual noise only marginally improves after \textit{uv}-plane subtraction of the central point source using the \texttt{CASA} \texttt{uvmodelfit} task.

\subsection{Circumnuclear Dust}
\label{app:dust}



The high-resolution ALMA MFS imaging modeled previously in Section~\ref{app:mfs} and shown in Figure~\ref{fig:mfs} were hoped to provide additional constraints on heretofore unconfirmed circumnuclear thermal dust emission. The Band 6 \texttt{imfit} residuals do not support any resolved excess in the inner 50 mas ($\sim$4 pc), so we estimated a flux density upper limit for any extended thermal source by shifting a circular ($R = 0\farcs05$) aperture randomly to 100 continuum-free regions of the residual X2c2 MFS image to sample the noise, taking the standard deviation $\delta S_\mathrm{ext} = 0.75$ mJy as an effective 1$\sigma$ limit \citep[for additional discussion of practical noise recovery in ALMA imaging, refer to][]{tsukui23}. The enclosed dust mass limit

\begin{equation}
    M_\mathrm{dust} < \frac{\delta S_\mathrm{ext} D_L^2}{\kappa_\nu B_\nu(T_\mathrm{dust})}
\end{equation}

\noindent assumes a standard dust absorption coefficient $\kappa_\mathrm{850\,\mu m}$ and emissivity index $\beta = 2$ in the opacity relationship $\kappa_\nu = \kappa_\mathrm{850\,\mu m} (\nu/352.7\,\mathrm{GHz})^\beta$ \citep{draine03}. For $T_\mathrm{dust} = 25$ K and a standard dust-to-gas mass ratio of $\sim$0.015 \citep{Sandstrom2013}, this modified black body $M_\mathrm{dust} < 1.1\times 10^4$ \msun\ would correspond to a total gas mass $M_\mathrm{gas} \lesssim 7.5\times10^5$ \msun. A hotter $T_\mathrm{dust} \sim 100$ K that follows the upper dust temperature distribution in ETGs \citep{kokusho19} would result in $M_\mathrm{dust} < 2.3\times 10^3$ \msun\ and $M_\mathrm{gas} < 1.5\times 10^5$ \msun. These $M_\mathrm{dust}$ upper limits are between 6$-$30$\times$ more constraining than is the SED fitting of mid-IR flux densities at much more coarse resolution \citep{baes10} and the lower $M_\mathrm{gas}$ limit for hotter dust abuts against \coone-derived upper limits presented in Section~\ref{sec:disc_em}. Higher-frequency ALMA continuum imaging would better probe thermal emission up the Rayleigh-Jeans tail even as the synchrotron nucleus intensity decreases as $\nu^{-0.6}$ (see Appendix~\ref{app:specslope}). However, DR limitations would dilute these improvements with nominal Bands 7 and 8 $\mathrm{DR}\sim 400$ and 250, respectively. Good observing conditions and standard self-calibration practices regularly allow for a factor of 2$\times$ higher DR \citep[][]{privon24} reaching up to $\sim$2500 in Band 7 \citep{komugi22}. Even in this best-case scenario, an estimated $\delta S_\mathrm{ext,\,Band\,7} \lesssim 0.5$ mJy limit would only probe 6$\times$ lower $M_\mathrm{dust}$ than above.

\section{Light Curves and Spectral Slopes}
\label{app:specslope}

\begin{figure}
    \centering
    \includegraphics[width=\textwidth]{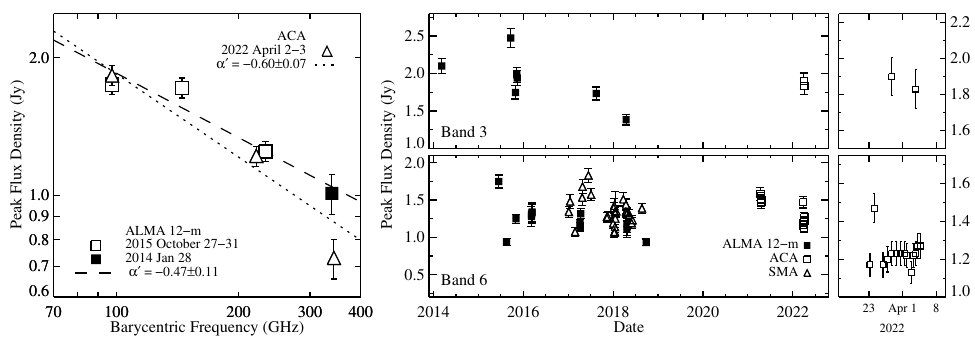}
    \caption{ALMA multi-Band peak flux densities of the M87 nucleus. The mm/sub-mm SED (\textit{left panel}) focuses on more contemporaneous points, showing separate power-law fits to the ALMA 12-m and 7-m data points. The 12-m data were taken far enough apart that the factor of $\sim$2$\times$ variability over few-month to year timescales (\textit{right panels}) somewhat dilutes its confidence. For the .V programs, peak flux densities were substituted with EHT ALMA-mode results \citep{event21c,event24}. Band 6 peak flux densities are consistent with the 1.3 mm light curve (spanning $\sim$1.2$-$2.3 Jy) measured at earlier epochs using the Submillimeter Array \citep[SMA;][]{bower15}.}
    \label{fig:cont_sed}
\end{figure}

In addition to ALMA 12-m observations taken between 2014 to 2019, we retrieved additional ACA 7-m imaging in projects 2019.1.00807.S and 2021.1.01398.S ($\overline{\theta}_\mathrm{FWHM} \sim 5\arcsec$ in Band 6), which monitored the Band 3 and 6 variability of M87 around EHT campaigns in 2021 and 2022. We obtained these archival data and applied standard pipeline calibration followed by phase-only self-calibration prior to MFS imaging with typical limiting sensitivities of 60--80 mJy beam\per. To construct more complete light curves shown in Figure~\ref{fig:cont_sed}, we joined together ALMA 12 and 7-m peak flux densities with Submillimeter Array (SMA) monitoring of M87 over this time frame \citep{event21c,event24}.

From adjacent ACA observations in program 2021.1.01398.S that span $\sim$100$-$400 GHz, we measured an apparent mm/sub-mm spectral index $\alpha^{\prime} = -0.60\pm0.07$ from roughly contemporaneous data taken between 2022 April 2$-$3. Fit errors include both rms and absolute flux scaling uncertainties. More separate ALMA 12-m imaging in programs 2012.1.00661.S and 2015.1.01352.S covering $\sim$95$-$340 GHz were obtained between 2014 Jan 28 and 2015 October 27$-$31 and result in a best-fit $\alpha^{\prime} = -0.47\pm0.11$. Both spectral indices are consistent with a predominately synchrotron origin for the radio-continuum data for the core and jet regions \citep[e.g.,][]{biretta91,event21c}. The more contemporaneous ALMA 12-m peak flux densities hint at a flattening spectral slope at longer wavelengths \citep{doi13}, although we do not detect a transition to $\alpha > 0$ for $\lambda \gtrsim 3$ mm that is seen in VLBI data \citep[][]{prieto16}.

\begin{deluxetable}{ccccccccc}[t]
\tabletypesize{\footnotesize}
\tablecaption{M87 ACA 7-m Observations and Imaging Properties\label{tbl:aca_cont}}
\tablewidth{0pt}
\tablehead{
 \multicolumn{2}{c}{Obs.} & \colhead{Member} & \colhead{$\overline{\nu}_\mathrm{obs}$} & \colhead{$\overline{\theta}_\mathrm{FWHM}$} & \colhead{Time} & \colhead{Peak $S_\nu$} & \colhead{MFS rms} & \colhead{PWV} \\[-1.5ex]
\multicolumn{2}{c}{Date} & \colhead{OUS} & \colhead{GHz} & \colhead{(arcsec)} & \colhead{(min)} & \colhead{(Jy)} & \colhead{(mJy)} & \colhead{(mm)} \\[-1.5ex]
\multicolumn{2}{c}{(1)} & \colhead{(2)} & \colhead{(3)} & \colhead{(4)} & \colhead{(5)} & \colhead{(6)} & \colhead{(7)} & \colhead{(8)}
}
\startdata
    \multicolumn{9}{c}{\textbf{2019.1.00807.S} (PI: Ramakrishnan)} \\ \cline{1-9} 
     & Apr 11 & A001/X1527/X2bc &     221.10 & 6\farcs94 & 7.55 & 1.58 & 83.1 & 1.62 \\[1ex]
     & Apr 15 & A001/X1527/X2bf &     221.10 & 6\farcs95 & 7.55 & 1.55 & 80.4 & 1.14 \\[1ex]
    2021 & Apr 17 & A001/X1527/X2c2 & 221.10 & 6\farcs96 & 7.55 & 1.52 & 79.7 & 0.69 \\[1ex]
     & Apr 18 & A001/X1527/X2c5 &     221.10 & 6\farcs77 & 7.55 & 1.50 & 80.3 & 0.81 \\[1ex]
     & Apr 19 & A001/X1527/X2c8 &     221.10 & 8\farcs65 & 7.55 & 1.47 & 85.8 & 0.81 \\[1ex] \cline{1-9}
    \multicolumn{9}{c}{\textbf{2021.1.01398.S} (PI: Ramakrishnan)} \\ \cline{1-9}
     & Mar 23 & A001/X15bc/X6a5 & 221.10 & 6\farcs05 & 7.55 & 1.17 & 59.1 & 1.00 \\[1ex]
     & Mar 24 & A001/X15bc/X6a8 & 221.10 & 4\farcs69 & 7.55 & 1.47 & 76.8 & 0.26 \\[1ex]
     & Mar 26 & A001/X15bc/X6ae & 221.10 & 6\farcs95 & 7.55 & 1.20 & 65.8 & 0.81 \\[1ex]
     & Mar 27 & A001/X15bc/X6b1 & 221.10 & 6\farcs03 & 5.02 & 1.23 & 67.4 & 0.76 \\[1ex]
     & Mar 28 & A001/X15bc/X7f8 & 97.50 & 14\farcs68 & 7.55 & 1.90 & 106.1 & 2.16 \\[1ex]
     & Mar 28 & A001/X15bc/X6b4 & 221.97 & 6\farcs34 & 7.55 & 1.23 & 62.2 & 2.11 \\[1ex]
     & Mar 29 & A001/X15bc/X6b7 & 221.10 & 6\farcs64 & 7.55 & 1.23 & 63.8 & 2.27 \\[1ex]
    \multirow{2}{*}{2022} & Mar 30 & A001/X15bc/X6ba & 221.10 & 6\farcs31 & 7.55 & 1.23 & 62.4 & 1.19 \\[1.5ex]
     & Mar 31 & A001/X15bc/X6bd & 221.10 & 6\farcs72 & 7.55 & 1.23 & 62.9 & 1.36 \\[1ex]
     & Apr 1 & A001/X15bc/X6c0 & 221.10 & 6\farcs76 & 7.55 & 1.22 & 62.0 & 0.30 \\[1ex]
     & Apr 2 & A001/X15bc/X26a & 203.00 & 6\farcs11 & 5.02 & 1.13 & 58.1 & 1.47 \\[1ex]
     & Apr 2 & A001/X15bc/X278 & 343.50 & 4\farcs19 & 5.84 & 0.73 & 79.6 & 0.80 \\[1ex]
     & Apr 2 & A001/X15bc/X6c3 & 221.10 & 6\farcs45 & 7.55 & 1.22 & 61.8 & 0.87 \\[1ex]
     & Apr 3 & A001/X15bc/X7fb & 95.70 & 16\farcs49 & 5.02 & 1.83 & 107.0 & 1.22 \\[1ex]
     & Apr 3 & A001/X15bc/X6c6 & 221.10 & 6\farcs72 & 7.55 & 1.23 & 63.2 & 1.36 \\[1ex]
     & Apr 4 & A001/X15bc/X271 & 233.00 & 7\farcs36 & 5.02 & 1.27 & 67.5 & 0.64 \\[1ex] \cline{1-9}
\enddata
\begin{singlespace}
  \tablecomments{Continuum properties of archival ACA 7-m data sets. Cols.\ (3) and (4) give the average frequency of the final MFS image across all spws and average synthesized beam FWHM from Briggs ($r=0.5$) imaging. Col.\ (5) gives the on-source integration time while cols.\ (6) and (7) report the peak and rms intensities of each MFS image. Col.\ (8) reports average PWV conditions. Each project used the same $\Delta\nu_\mathrm{obs} = 15.6$ MHz TDM channel binning with dual polarization set-up.}
\end{singlespace}
\end{deluxetable}

At radio to mm wavelengths, the M87 nucleus light curve has been (visually) consistent with a damped-random walk (DRW) over the past decade or so with $\sim$50\% variability on $\sim$few month timescales \citep[Figure~\ref{fig:cont_sed};][]{acciari09,hess11,hada14,bower15} preceded by a larger flaring event in 2005 spanning a few years \citep{abramowski12}. Strong, persistent variability may introduce large uncertainties in the SED slope measured from non-contemporaneous data across a relatively narrow frequency range. We tested the impact for the above $\alpha^\prime$ by creating mock light curves in ALMA Bands 3, 6, and 7 whose variability was defined by the power spectral density (PSD) fits to other radio galaxies \citep[from 43 GHz light curves with an average power-law PSD slope $\alpha_\mathrm{PSD} = -1.4$;][]{zhang20}. Following the methodology outlined by \citet{emmanoulopoulos13}, we simulated an ALMA Band 6 peak light curve spanning 2 yrs with a one-day cadence. Additional Bands 3 and 7 light curves were then created using a consistent (intrinsic) $\alpha_\mathrm{intr} = -0.6$ at each time epoch. We evaluated the change in recovered $\alpha^{\prime}$ relative to $\alpha_\mathrm{intr}$ when the Bands 3 and 7 measurements were drawn from days to months separate from the Band 6 values. For each synthetic measurement, we added random, realistic noise drawn from a normal distribution following standard absolute flux calibration uncertainties \citep{fomalont14}. We then collected these $\alpha^{\prime}$ for a given time separation across the entire synthetic light curve and take the standard deviation of the $\alpha_\mathrm{intr} - \alpha^{\prime}$ values as a proxy for the total uncertainty $\delta\alpha^{\prime}$ when using non-contemporaneous ALMA measurements. When synthetic measurements are separated by at most a few days, we find a standard deviation $\delta \alpha^{\prime}\sim 0.05$; for few-week to few-month separations, $\delta\alpha^{\prime}$ increases from $\sim$0.15 until flattening around $\sim$0.35. As a result, the ALMA 12-m $\alpha^\prime = -0.47\pm0.11$ that includes an $S_\mathrm{\nu,Band\,7}$ point taken over 1.5 years before the remaining closely-sampled Bands 3--6 data underestimates the true uncertainty by at least a factor of 3. We note that $\alpha^{\prime}$ measured across broad radio (or radio-to-mm) frequency ranges will be less susceptible to uncertainty introduced by DRW variability.

\section{Peak Continuum Spectra}
\label{app:peak}

In Figure~\ref{fig:contabs}, we present peak flux density spectra of all member OUS listed in Table~\ref{tbl:sample}. The spectra show no strong molecular line features except for residual contamination from atmospheric ozone, which we discuss in Appendix~\ref{app:atm}. Spectral shapes are not always consistent across all spws in a single setup or as a function of time. We attribute this to either limitations in the standard pipeline calibration process or assumptions about the calibrator spectral index.

ALMA bandpass and flux calibrations typically rely on week to few-month cadence monitoring of bright AGN ($S_\nu\sim0.1-2$ Jy) that are tied to Solar System standards \citep{bonato19}. For point-source AGN, mm-wavelength flux densities have a median $\sim$10\% variability amplitude over $\sim$year timescales \citep[and larger at higher frequencies; e.g.,][]{sadler08,bonato19}. Calibrator flux densities are updated in the ALMA Calibrator Source Catalogue (CSC) and recorded in pipeline files for MS reconstruction. This approach does not always account for the power-law spectral index of a calibrator (reflected by the \texttt{spix} parameter in the \texttt{setjy} task) or \texttt{spix} variability. Bandpass and flux calibrators are overwhelmingly selected from radio quasars and BL Lac-type objects that show approximately flat spectral shapes as measured in L to C radio bands. In ALMA Bands, however, the mm/sub-mm spectral index $\alpha$ is often more characteristic of classical synchrotron radiation, \citep[including for some calibrators used in these programs; see Appendix~\ref{app:specslope} and][]{bonato18}. In some cases, the pipeline calibration process adopted a non-zero \texttt{spix} based on a roughly contemporaneous spectral index measurement, but this was not uniform in the earlier ALMA Cycles.

\begin{figure}[!ht]
    \centering
    \includegraphics[width=0.95\textwidth]{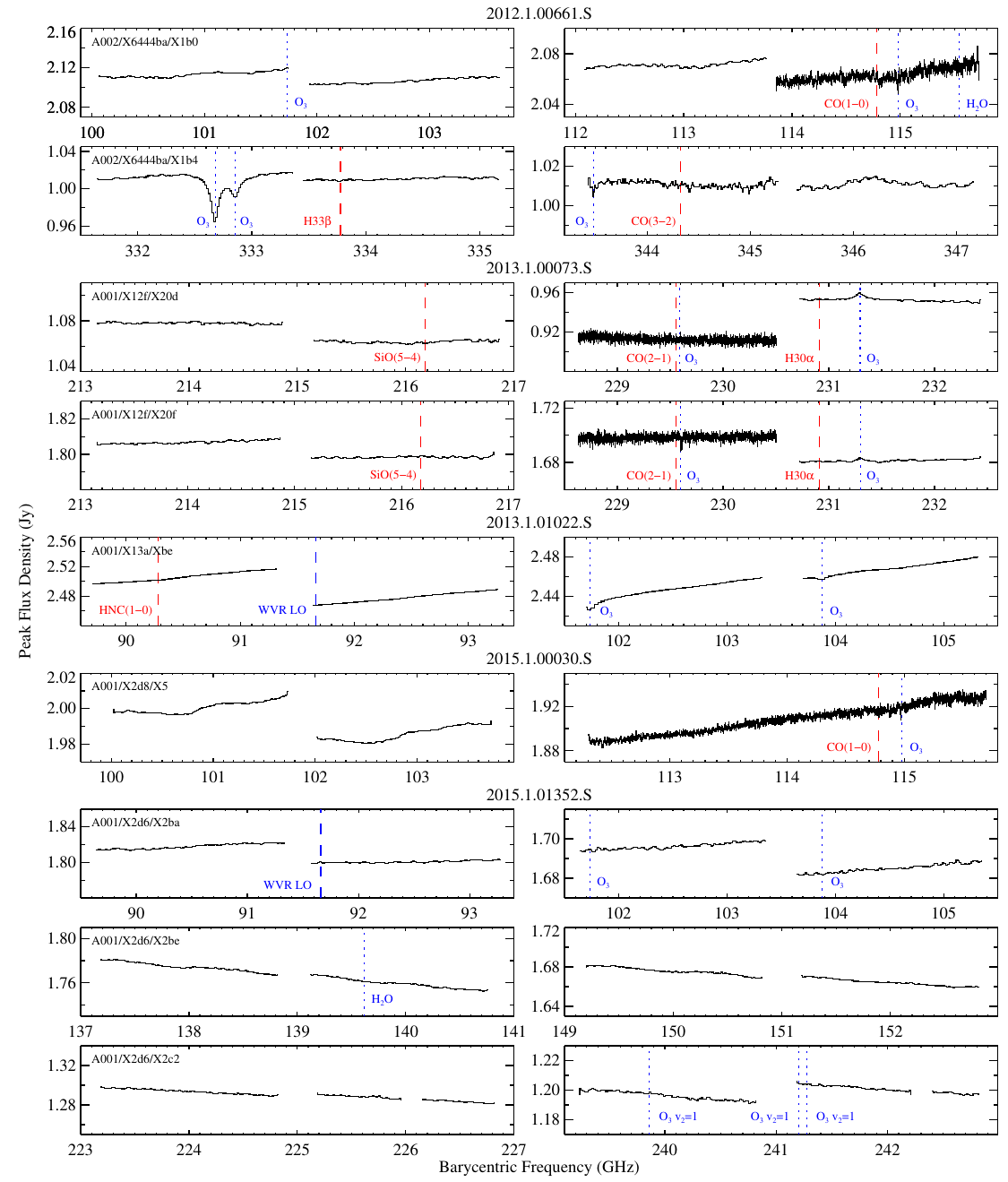}
    \caption{Peak flux density spectra of the M87 nucleus at the native channel binning after standard pipeline and self-calibration processes. Flux calibration was adjusted for overlapping spws in programs 2015.1.00030.S and 2016.1.00021.S only to ensure continuous peak flux densities. The frequencies of atmospheric ozone and water vapor transitions (vertical dotted blue lines; shifted to the barycentric frame), harmonics of the WVR local oscillator \citep[LO;][]{cortes24}, and extragalactic molecular gas and hydrogen recombination transitions (dashed red lines; redshifted by $z_\mathrm{obs}=0.004283$) are also shown. The characteristic scalloped pattern in the .V panels is an artifact of phasing the 12-m antennae in APP mode.}
    \label{fig:contabs}
\end{figure}

\addtocounter{figure}{-1}
\begin{figure}[!ht]
    \centering
    \includegraphics[width=0.92\textwidth]{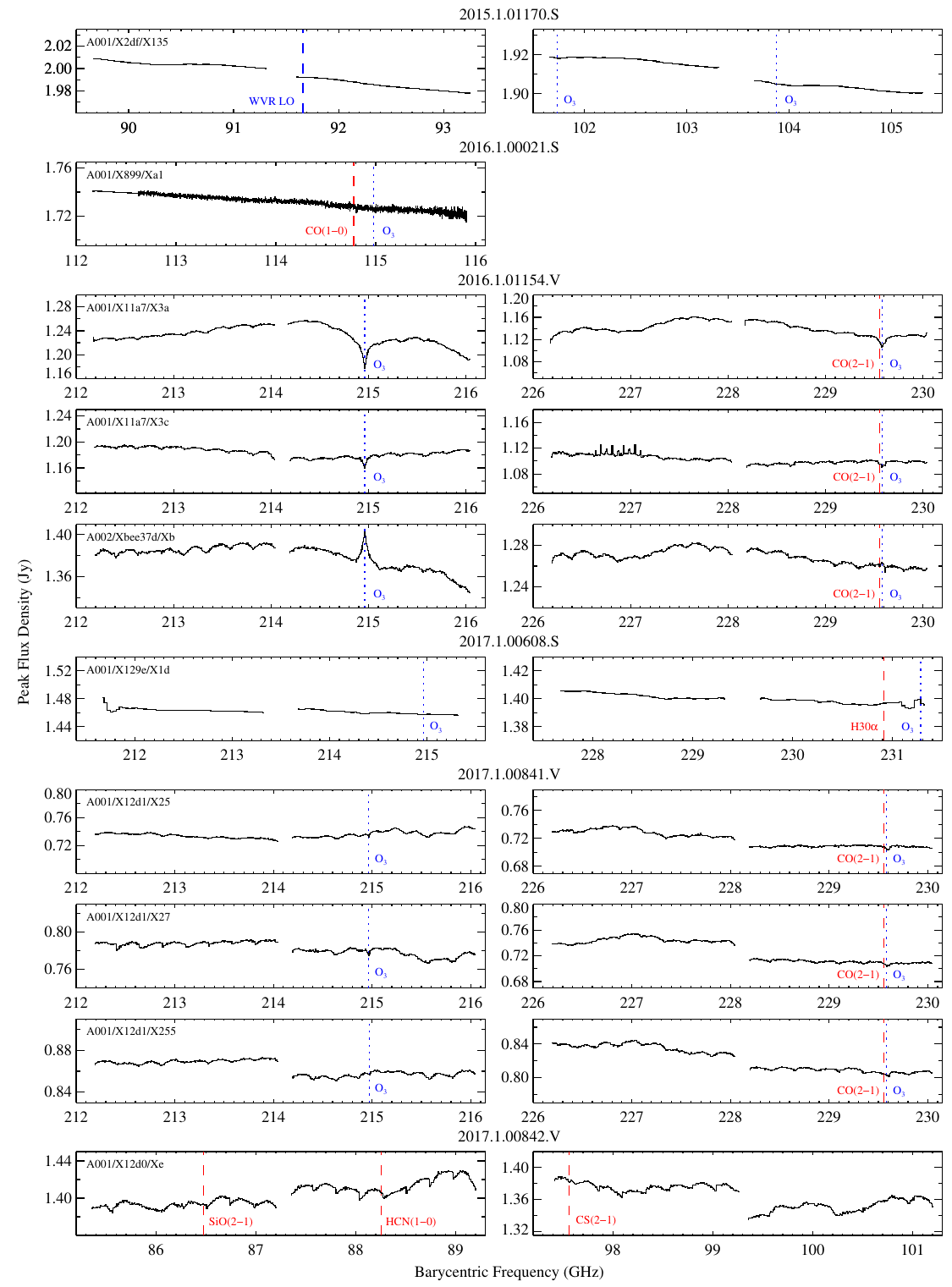}
    \caption{cont.}
\end{figure}

Within a single sideband, \texttt{spix} discrepancies led to inconsistencies in the relative M87 flux densities in individual spws at the 1--2\% level. For adjacent EBs separated by just an hour, even when using the same calibrators, the inter-spw $\alpha$ value changes by up to $\sim$10\% (see Figure~\ref{fig:bandpass}). The main outlier is the X20d data set, whose inter-spw $\alpha$ is much steeper than roughly contemporaneous ALMA Band 6 observations and SED slopes in Figure~\ref{fig:cont_sed}. We attribute this discrepancy to poor spw-dependent phase solutions during self-calibration. Over all spws for a given spectral setup, the typical inter-spw $\alpha$ after standard pipeline and self-calibration processes spans a broad range of $-$0.1 to $-$1.0. Some of this scatter for Band 3 data may be attributable to the apparent $S_\nu$ flattening for $\nu \lesssim 100$ GHz \citep[see Figure~\ref{fig:cont_sed} and][]{prieto16}. Intra-spw slopes are particularly unreliable \citep{francis20a}, in part due to bandpass calibration limitations (see Appendix~\ref{app:bandpass}).

\section{Atmospheric Modeling}
\label{app:atm}

\begin{figure*}[!t]
    \gridline{\fig{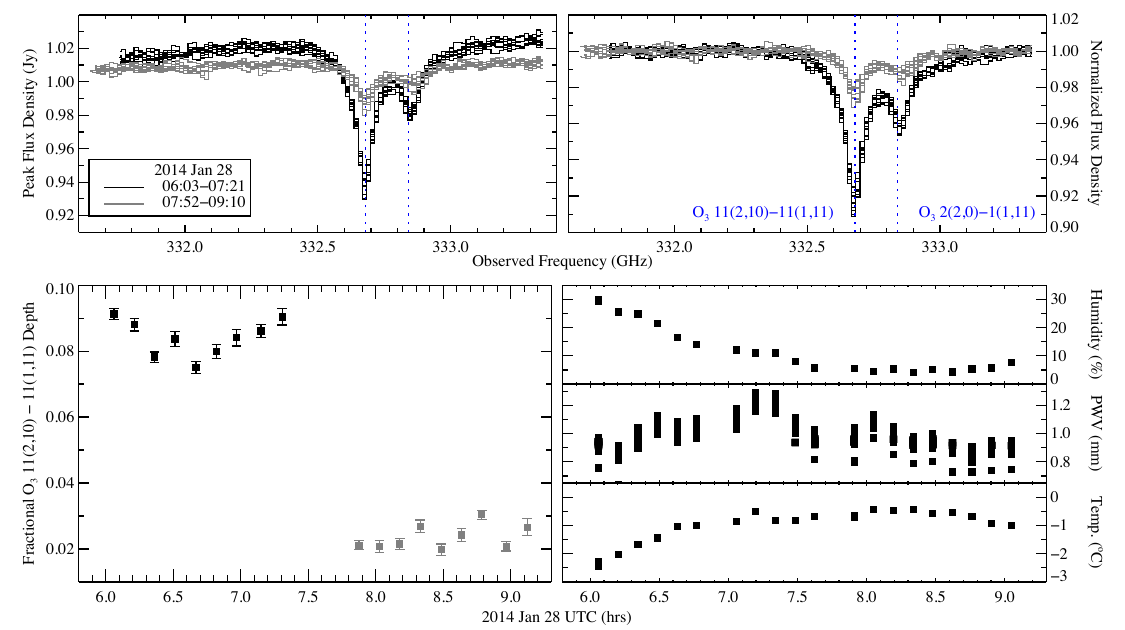}{0.9\textwidth}{}}
    \gridline{\fig{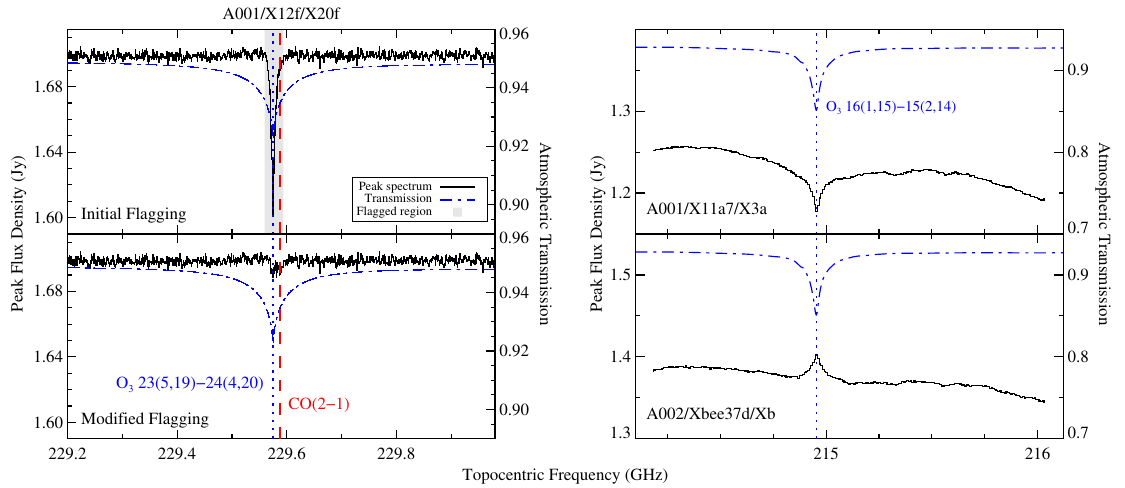}{0.9\textwidth}{}}
    \caption{M87 peak continuum spectra showing the residual impact of ozone after pipeline and self-calibration processes. In the Band 7 data (\textit{top and middle panels}), peak spectra imaged by scan show a decrease in residual O$_3$ line depth (\textit{middle left}; CDMS/JPL line intensities $-$3.66 to $-$4.08) after changing to a more local bandpass calibrator between EBs. Over this time frame, moderate changes in relative humidity and PWV (about the average; dashed line) were detected while the ground temperature remained relatively stable (\textit{middle right}). In Band 6 (\textit{bottom panels}), some O$_3$ features are not fully corrected by pipeline calibration. For the X20f data set (\textit{bottom left}), \texttt{CASA} calibration using an earlier \texttt{ATM} version showed a narrow absorption feature coincident with an O$_3$ line (CDMS/JPL intensity $-$4.61) whose core was manually flagged (shaded region) in the bandpass calibrator as an extragalactic line while the broad wings were largely corrected. After removing these flags, the standard calibration script corrected for nearly all the narrow O$_3$ impact, revealing the faint, narrow \cotwo\ absorption. Some residual O$_3$ wing excess remains due to calculating $T_\mathrm{sys}$ with TDM binning \citep{tadhunter08}. For peak spectra from program 2016.1.01154.V covering a prominent O$_3$ line (\textit{bottom right}; CDMS/JPL intensity $-$4.19), two EBs observed nearly a week apart show telluric line undercorrection (\textit{top}) and overcorrection (\textit{bottom}) at the 25$-$50\% level during pipeline calibration. For the Band 6 cases, the \texttt{am} atmospheric transmission model \citep{paine18} is overlain for comparison.}
\label{fig:ozone_var}
\end{figure*}

The deepest atmospheric absorption features (primarily water vapor) often set the frequency ranges for mm/sub-mm Bands. For Band 3, an O$_2$ line ($N=1-1$ $J=1-0$; $\nu_\mathrm{sky} = 118.750$ GHz) limits coverage in practice to a topocentric $\nu_\mathrm{sky} < 115.93$ GHz. As we demonstrate in Appendix~\ref{app:bandpass}, this impacts the continuum subtraction reliability near \coone\ and may contribute to the spurious broad absorption or emission-line features. Pipeline calibration does correct for most of the transmission drop due to broad H$_2$O and O$_2$ absorption. However, in the X1b4 peak continuum spectrum the spw nearest the broad H$_2$O feature centered at $\nu_\mathrm{sky} = 325.15$ GHz shows some residual spectral curvature in the first EB (see Figure~\ref{fig:ozone_var}) as the relative humidity decreases rapidly after the initial calibration scans.

From the continuum spectral cubes, we identified several absorption and emission features that could not be matched to Galactic or extragalactic lines but that did coincide with various telluric ozone (O$_3$) transitions in the topocentric frame. \citet{hunter18} demonstrated that the coarse TDM atmospheric calibration used for these ALMA 12-m data may over/undercorrect for atmospheric ozone, resulting in either positive/negative residuals, respectively. These lines have been further studied by the PHANGS-ALMA team in Band 6 around redshifted \cotwo\ in their total-power array imaging, primarily as a function of elevation or airmass \citep{usero19,leroy21} while assuming that atmospheric conditions (e.g., relative humidity, ambient temperature, and O$_3$ concentration) are stable over 1$-$2 EBs (1$-$2 hours). Here, we focus on atmospheric variability on these timescales as well as the impact of calibrator offset on the correction of telluric line features. Since the O$_3$ transitions do not generally interfere with most of the extragalactic line transitions at the redshift of M87, we did not attempt to mitigate their impact by constructing $T_\mathrm{sys}$ spectra in FDM mode during manual calibration in \texttt{CASA} \citep[as suggested by][]{hunter18}. In Figure~\ref{fig:contabs} and elsewhere, we indicate the frequencies of the ozone and water vapor transitions that have CDMS/JPL logarithmic intensities\footnote{Identified using the Splatalogue database for astronomical spectroscopy \url{https://splatalogue.online/}} \citep{pickett98,muller05} brighter than $-$5.3, the limiting depth down to which ozone lines have been detected in these ALMA 12-m data. Fractional depths are only a fraction of a percent in the highest-SDR data sets.

To aid in identifying and characterizing faint ozone features, we employed the \texttt{am} atmospheric modeling tool \citep{paine18} to calculate an atmospheric transmission curve for each continuum spectrum. This approach adopts column densities for select molecules and approximates the vertical structure with layers of varying pressure and temperature. Without detailed conditions above the Atacama plateau, we borrowed the realistic \texttt{am} manual model for Mauna Kea at nearly the same altitude after adjusting for the average PWV, target elevation, and ground-layer temperature for each MS. Using the same spectral ranges and binning, we constructed final transmission curves for each spw and shifted them to barycentric frame to more easily compare to $S_\nu$ plots in Figure~\ref{fig:contabs}. These \texttt{am} model results are close to those from the \texttt{Atmospheric Transmission at Microwaves} model \citep[\texttt{ATM};][]{pardo01,pardo19} used during \texttt{CASA} calibration, especially after \texttt{ATM} included additional minor lines and was tuned for both observed seasonal and diurnal O$_3$ concentration and altitude changes \citep{frith20,pardo24}.

These ozone features are typically distinct from major molecular transitions at a barycentric $z_\mathrm{obs} = 0.004283$, although one O$_3$ transition [23(5,19)$-$24(4,20); $\nu_\mathrm{sky} = 229.575$ GHz] lies close to the \cotwo\ frequency for low redshift targets. For the X20f data set, the archival calibration script manually flagged a narrow range of channels for the bandpass calibrator (J1229+0203; 3C 273), misidentifying an atmospheric ozone line as extragalactic in origin. The result was a sharp, moderately deep \citep[$\sim$6\% of the continuum level; see Figure~\ref{fig:ozone_var} and][]{ray24}, spurious absorption feature in the corresponding calibrated M87 spectral window near redshifted $\nu_\mathrm{CO(2-1)}$. During pipeline calibration, the wings of this ozone feature remain apparent as slight excesses due to $T_\mathrm{sys}$ calculations being carried out in TDM mode. After inspecting the relevant frequency range when this calibrator was a science target in project 2019.1.00807.S, we could not identify a line feature or possible molecular gas candidate at the calibrator's redshift ($z_\mathrm{obs} = 0.15834$). We removed these flags before re-calibrating the raw data and re-imaging the X20f MS. This removed nearly all of the ozone line contamination and revealed the faint \cotwo\ absorption feature. None of the other data sets appear to have similar flagging issues.

Atmospheric phase differences between baselines vary with characteristic timescales of minutes. While many studies of water vapor or O$_3$ column densities tend to track differences on hour timescales or longer \citep[e.g.,][]{rappengluck14,zhang14,maud23}, rapid changes decrease the coherence time for high-frequency observations \citep{matsushita17} and increase phase fluctuations, especially for extended ALMA configurations \citep{privon24}. Regular WVR measurements correct for smaller-scale, more rapidly-varying PWV changes, and phase self-calibration works to remove residual path length discrepancies. Due to the bright mm/sub-mm continuum source at the center of M87, clean phase solutions can generally be obtained through iterative self-calibration by using the \texttt{CASA} \texttt{gaincal} task on timescales as short as the scan length. We find that these phase self-calibration loops mitigate nearly all path length fluctuations that arise due to tropospheric turbulence, even on the longest ALMA baselines. However, for two exceptions -- the X20d data set and the first EB of X1b0 (see Figure~\ref{fig:colineabsfit}) -- the standard self-calibration process never produced sufficiently clean phase solutions regardless of the choice of \texttt{refant} or \texttt{solint}. The resulting spectra were less reliable due to poorer rms sensitivity or much lower flux scaling as well as high \texttt{CLEAN} residuals.

Good bandpass calibrators are not always found close to the science target on the sky, but those that are significantly removed in time or obtained along a far-removed line of sight have the potential to leave imprints of atmospheric lines on calibrated science data \citep{meena10,frith20}. In Figure~\ref{fig:ozone_var}, we show peak $S_\mathrm{\nu,Band\,7}$ spectra obtained in the X1b4 data set during Cycle 1, showing residual O$_3$ absorption [in the 11(2,10)$-$11(1,11) and 2(2,0)$-$1(1,1) transitions; $\nu_\mathrm{sky} = 332.705$ and 332.882 GHz, respectively]. The \texttt{am} model with matching observing conditions shows a fractional depth of $\sim$0.18 for the deeper O$_3$ line compared to the neighboring continuum. In this member OUS, the bandpass calibrators for the first (J1058+0133) and second (J1229+0203) EB were separated by 25.3\degr\ and 10.3\degr\ from M87, respectively, and \texttt{ATM} models removed about 55\% and 90\% of the expected ozone absorption towards M87 during the standard \texttt{CASA} pipeline calibration. For the second EB, the bandpass calibrator appears to have been observed through a more similar $N_\mathrm{O_3}$, enabling a better telluric correction. We note that PWV or relative humidity trends are not mirrored by the residual O$_3$ line strengths as the molecules originate from different altitudes (O$_3$ primarily at $\sim$15$-$40 km while the H$_2$O scale height is 1$-$2 km). Variable $N_\mathrm{O_3}$ is suggested by other data sets with both distant and close bandpass calibrators, especially in projects 2013.1.00073.S and 2013.1.01022.S (both using J1229+0203), and the very strong residuals in 2016.1.01154.V (using B1730-130 and 4C 01 28, with separations of 79.1\degr\ and 25.3\degr, respectively). Positive O$_3$ line features in the target spectral cube suggest a higher $N_\mathrm{O_3}$ towards the bandpass calibrator than towards the science target. Assuming that $N_\mathrm{O_3}$ scales linearly with the depth of unsaturated O$_3$ absorption lines, the residual depths shown in Figure~\ref{fig:ozone_var} for a single EB suggest that either stratospheric or ground-layer $N_\mathrm{O_3}$ along the line of sight changed by up to $\sim$6\% over just 30 min.

\section{Channel Maps}
\label{app:channel}

\begin{figure*}[!ht]
    \gridline{\fig{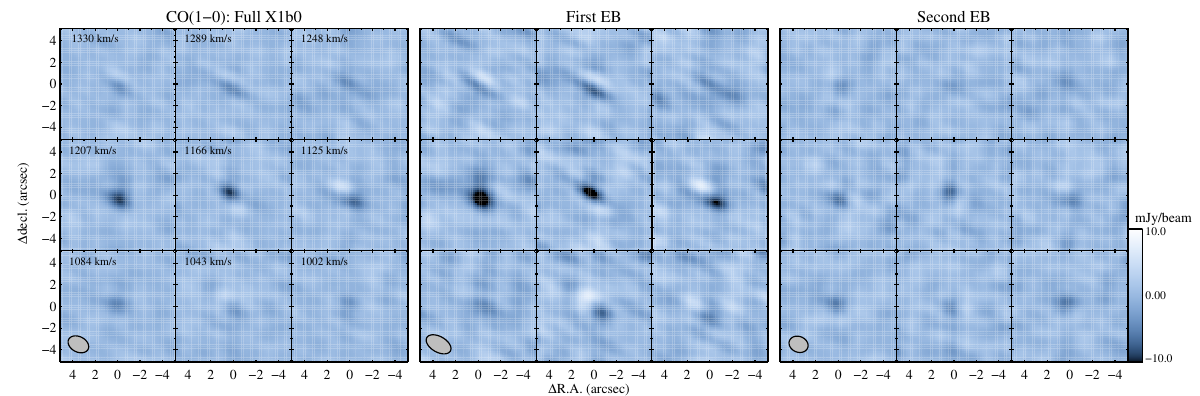}{\textwidth}{}}
    \gridline{\fig{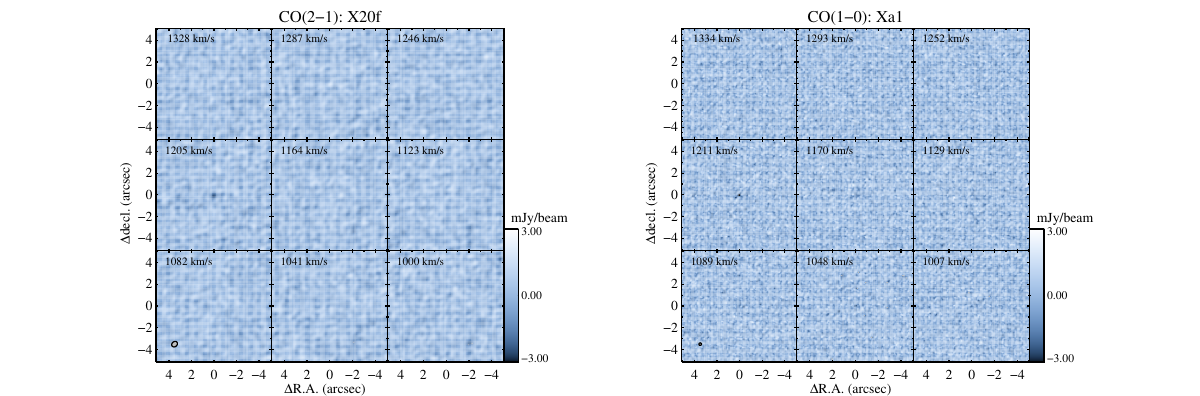}{\textwidth}{}}
    \caption{CO channel maps covering a $10\arcsec \times 10\arcsec$ centered on the M87 nucleus for the X1b0, X20f, and Xa1 data sets. These ALMA 12-m data were imaged into $\sim$40 \kms\ channels to improve S/N and to compare with the channel maps of \citet{ray24}. Channel velocities are listed and synthesized beams are shown as ellipses (\textit{bottom left panels}). X1b0 channel maps (\textit{top}; $\overline{\theta}_\mathrm{FWHM} \sim 1.5\arcsec-2\arcsec$) were taken from continuum-subtracted cubes using the full MS as well as using only the first or second EB. The first EB has poorer phase solutions, leading to worse continuum subtraction and more spurious features about the removed nuclear continuum than the second EB. Slight oversubtraction of the point-like nuclear continuum is visible around the redshifted \coone\ transition. In the first EB cube, the apparent excess features show no coherent velocity behavior. Channel maps from the X20f  (\textit{bottom left}; $\overline{\theta}_\mathrm{FWHM} \sim 0\farcs 51$) and Xa1 cubes (\textit{bottom right}; $\overline{\theta}_\mathrm{FWHM} \sim 0\farcs 21$) do not show any of the possible CO emission, although they do show faint central CO absorption at the appropriate channel (1205$-$1210 \kms).}
\label{fig:comp_channels}
\end{figure*}

While most CO emission in other nearby cool-core galaxies is resolved on sub-arcsecond scales, the Xa1 data set with MRS~$\gtrsim 3\arcsec$ cannot entirely rule out diffuse emission spread out over a couple of the larger regions with $\sim$5\arcsec\ extent. Line profiles shown in Figure~\ref{fig:lineprof} do not show any strong support for \coone\ emission in the CND or the other regions. However, \citet{ray24} purport to find evidence for a CND traced by \coone\ with a $\sim$200-pc extent. Their work relies on the X1b0 data set ($\overline{\theta}_\mathrm{FWHM} \sim 1\farcs 7$; MRS~$\sim 12\arcsec$) from ALMA Cycle 1 with $\sim$40 \kms\ channel binning. In Figure~\ref{fig:comp_channels}, we present circumnuclear channel maps from continuum-subtracted cubes in the X1b0, Xa1, and X20f data sets with similarly broad binning. For the X1b0 data set, we include maps made from either of the two adjacent EBs. We find that phase self-calibration and continuum subtraction for the first EB (rms = 0.95 mJy beam\per\ per channel) was less successful at removing the continuum contributions than for the second EB (rms = 0.42 mJy beam\per), resulting in sidelobe features at the $\sim$2\% level that appear around the removed central source (see Figure~\ref{fig:contabs}). Channel maps for the second EB of the X1b0 data set as well as the Xa1 and X20f data show no clear \coone\ or \cotwo\ sources near the center of M87.

In the case of unresolved CO emission, the H$_2$ surface mass density in \msun\ pc\pertwo\ over a single beam area is:

\begin{equation}
    \Sigma_\mathrm{H_2,\,limit} \approx 1220 \left[\frac{\alpha_\mathrm{CO}}{\nu^2 \theta_\mathrm{maj}\theta_\mathrm{min}}\right]\Delta v \times \mathrm{rms} \,,
\end{equation}

\noindent where $\nu$ is in GHz, the synthesized beam FWHM are in arcseconds, the point-source sensitivity rms is in mJy, and the channel width $\Delta v$ is in \kms. Using the same $\alpha_\mathrm{CO(1-0)}$ as in Section~\ref{sec:narrowem}, the composite X1b0 data (rms = 0.57 mJy beam\per\ per channel) should give a $\Sigma_\mathrm{H_2,\,limit} \sim 2.3$ \msun\ pc\pertwo\ in each beam. Over the putative $R\sim 200$ pc extent for a slightly flattened CND, this $\Sigma_\mathrm{H_2}$ 1$\sigma$ limit would translate to $M_\mathrm{H_2,\,limit}\sim 2\times 10^5$ \msun, similar to the limit from non-detection in the more physically-motivated CND or annulus regions in Table~\ref{tbl:narrowem}. Near the redshifted CO transition, the Xa1 data (rms = 0.34 mJy beam\per\ per channel; $\overline{\theta}_\mathrm{FWHM} \sim 0\farcs 21$) should be sensitive to $\sim$84 \msun\ pc\pertwo\ while the X20f data (rms = 0.44 mJy beam\per; $\overline{\theta}_\mathrm{FWHM} \sim 0\farcs 49$) and an $R_{21} \sim 0.7$ ratio of \cotwo\ to \coone\ luminosities gets down to $\sim$7.5 \msun\ pc\pertwo.

\section{Constraints on Broad CO Features}
\label{app:constraints}

Here, we present broad but very low-amplitude spectral deficits or excesses about redshifted CO when excluding wide channel ranges during continuum subtraction. To aid in future line studies, we detail evidence that they arise naturally from the inherent limitations of ALMA bandpass calibration instead of broad CO absorption or emission.

\subsection{Broad CO Absorption}
\label{app:broadabs}

\begin{figure*}[!t]
    \centering
    \includegraphics{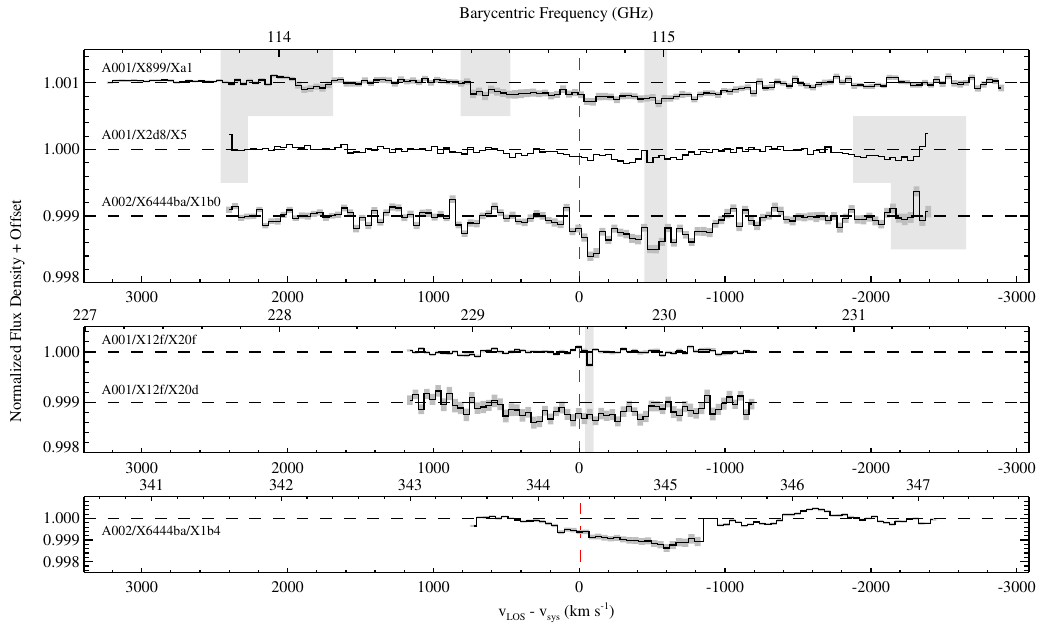}
    \caption{Continuum-subtracted spectra of the M87 nucleus integrated in a circular aperture ($R\sim \overline{\theta}_\mathrm{FWHM}$). To these, we added in a smooth fit to the peak continuum outside of $|\vlos - \vsys| < 1000$ \kms\ before normalizing. Spectra covering the \coone\ (\textit{top panel}), \cotwo\ (\textit{middle}), and \cothree\ transitions (\textit{bottom}) were imaged using coarse (40 \kms) binning. Shaded regions indicate uncertain flux densities, due to either residual contamination by narrow atmospheric lines or bandpass calibration issues at spw edges for overlapping spws. The apparent Band 3 spectral deficits between about $-1000 < \vlos-\vsys < 100$ \kms\ could be interpreted as a very tenuous, fast molecular gas outflow traced only by \coone\ absorption. However, inconsistent Band 6 and 7 results as well as bandpass calibration limitations (discussed in Appendix~\ref{app:bandpass}) make that scenario implausible.}
    \label{fig:broadabs}
\end{figure*}

The $\pm$500 \kms\ range used before to explore CO emission or absorption may be too narrow to study possible molecular outflows \citep[e.g.,][]{cicone14}. After excluding channels with $|\vlos - \vsys| < 1000$ \kms\ in the \texttt{uvcontsub} task, we imaged the deeper X1b0, X5, and Xa1 data sets (Band 3) as well as those from X20d and X20f (Band 6) covering redshifted \coone\ and \cotwo\ into spectral cubes with $\sim$40 \kms\ channel binning. For the X1b4 data set (Band 7), we excluded a narrower channel range with $-1000 < \vlos - \vsys < 400$ \kms\ due to more limited frequency coverage when constructing a \cothree\ cube. Line profiles integrated over their central beam areas (Figure~\ref{fig:broadabs}) show apparent deficits in the $-1000 < \vlos - \vsys < 100$ \kms\ range. In Band 3, these broad spectral deficits translate to apparent peak opacities $\tau_\mathrm{0,CO(1-0)} \sim 0.002 - 0.004$ with S/N~$\sim 5-25$ relative to the rms of the continuum-fitting channels. Over that velocity range, the corresponding integrated $\tau_\mathrm{CO(1-0)} \sim 0.1-0.3$ \kms\ with S/N~$> 10$ would far exceed the narrow CO absorption in Table~\ref{tbl:narrowabs}. This apparent \coone\ opacity signature is not replicated in the higher-frequency Bands.

If real, these deficits would seem to trace a molecular outflow that is much faster than the blueshifted atomic outflows suggested by optical emission-line studies \citep{osorno23} or narrow absorption \citep{tsvetanov99b,sabra03}. Molecular ouflows have been detected in absorption against ultraluminous infrared galaxy nuclei and AGN at a few $\times$100 \kms\ to $\sim$1000 \kms\ \citep[including some LINERs;][]{veilleux13,stone16,nagai19,herrera20}. Such molecular outflows are also detected in emission with blueshifts reaching similar terminal speeds \citep[e.g.,][]{cicone14,garciaburillo14,riffel20,bolatto2021}. In LINERs like M87, however, the core luminosity is well below the expected threshold to drive a broad outflow \citep{veilleux13,prieto16}. The primary arguments against a fast-outflow interpretation are the inconsistent spectral behavior and the low amplitude. First, the velocity extremes vary by up to $\sim$600 \kms\ for the \coone\ transition. Second, peak fractional deficits $1-\exp^{-\tau_0} \approx 0.00025$ are at or below the level of the bandpass stability.

\subsection{Broad CO Emission}
\label{app:broademission}

\begin{figure*}[!t]
    \centering
    \includegraphics[width=0.95\textwidth]{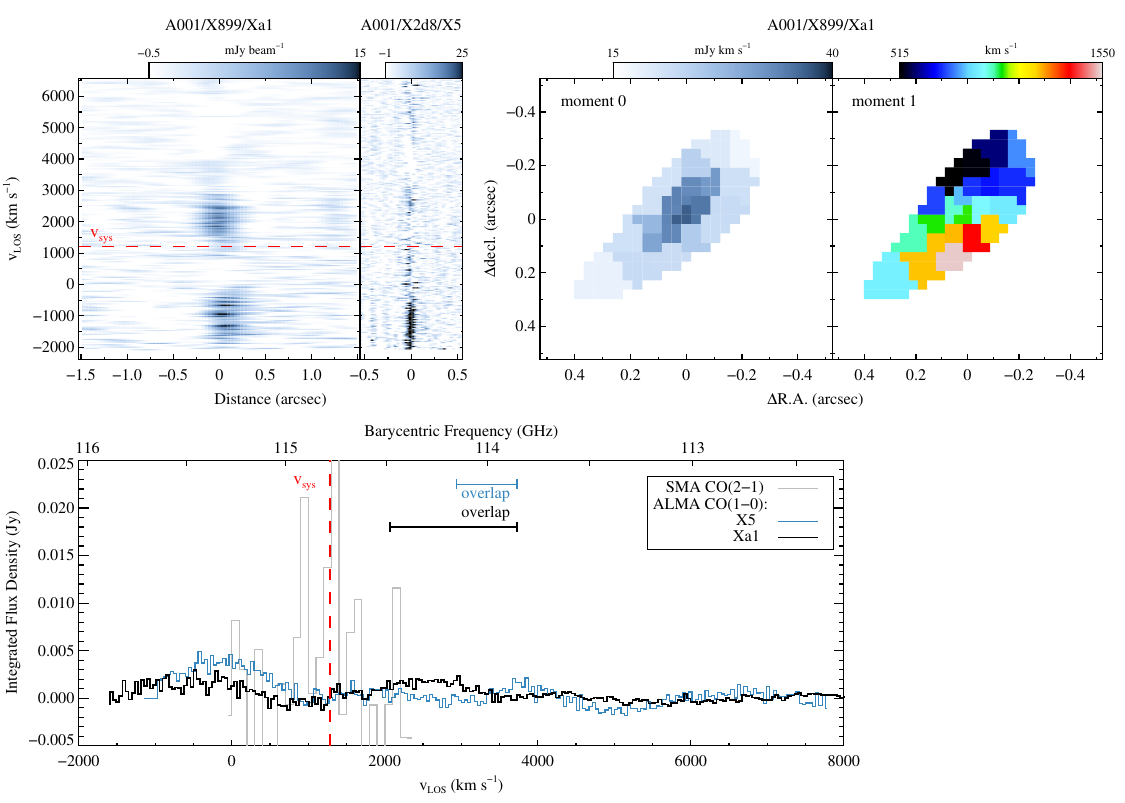}
    \caption{Spectral excess properties from continuum-subtracted M87 Xa1 and X5 data sets that contain the redshifted \coone\ transition (\vlos; dashed lines). Distinct, asymmetric line profiles integrated over a circular aperture ($R=\overline{\theta}_\mathrm{FWHM}$; \textit{bottom panel}) show larger spectral variation than the channel-to-channel noise ($\lesssim$0.05\% of peak continuum) measured at lower $\nu$. For comparison, we show the narrower SMA \cotwo\ line profile \citep[extracted using \texttt{PlotDigitizer} from][]{tan08}. Primary spw overlap regions are indicated. In large part because of the choice of continuum window, the PVDs extracted along the N-S direction (\textit{top left}) show essentially unresolved spectral excess between about $-2500 < \vlos - \vsys < -500$ \kms\ in both cases and between $200 < \vlos - \vsys < 2000$ \kms\ for the Xa1 data set. The excess/continuum flux ratios of 0.1$-$0.3\% for these two member OUS match the bandpass stability level. The zeroth and first moment maps from the Xa1 data set (\textit{top right}) are inconsistent with the atomic gas CND orientation and extent as well as the unresolved nature of the continuum feature seen in the higher-resolution X5 PVD. We therefore consider these features to be residual structures not fully removed during calibration.}
    \label{fig:emission}
\end{figure*}

When first exploring continuum subtraction and CO line profiles for these M87 data sets, we noticed hints of CO excess reaching out to $\pm$2000 \kms, which could be attributed to a pc-scale disky structure. To explore this possibility, we excluded channels with $100 < |\vlos - \vsys| < 2200$ \kms\ in the \texttt{uvcontsub} task (with \texttt{fitorder} = 1) and imaged redshifted \coone\ in the X5 and Xa1 data sets into cubes with 40 \kms\ bins. Care was taken to avoid channels at spw edges that showed poor response or whose flux densities were discrepant with those in overlapping spws. Unfortunately, higher-$J$ transitions did not have sufficient frequency coverage (or spw overlap) to allow for reliable continuum subtraction following these $|\vlos-\vsys|$ exclusions. In Figure~\ref{fig:emission}, we show \coone\ position-velocity diagrams (PVDs) extracted in a N--S direction with a width matching the beam average FWHM. The putative excesses are roughly centered around $\vlos - \vsys = -2200$ and +1000 \kms, with spatial centroids separated by as little as 0\farcs015 ($\sim$1 pc). Because this excess appeared to be almost unresolved in either case, we constructed line profiles by integrating over the central region with a circular $R=\overline{\theta}_\mathrm{FWHM}$ aperture.

From the more symmetrical Xa1 CO line profile, the integrated excess of 4.5 Jy \kms\ would be contained within the X5 beam ($R<4$ pc) and correspond to $M_\mathrm{H_2} \sim 3\times 10^6$ \msun\ and $\overline{\Sigma}_\mathrm{H_2} \gtrsim 2\times 10^4$ \msun\ pc\pertwo, far exceeding central molecular gas surface mass densities on similar scales in similar targets \citep{garciaburillo16}. A measured \coone\ map for the Xa1 data set is inconsistent with the ionized gas disk orientation; also, the putative \coone\ emission is much smaller than $\overline{\theta}_\mathrm{FWHM}$ for the Xa1 data set, so a rotating disk cannot explain the moment 1 map gradient in the N--S direction\footnote{We tentatively detect a similar small shift in continuum centroid (but not peak location) across a spw in both the X5 and Xa1 continuum cubes shown in Figure~\ref{fig:contabs}. We attribute this to residual frequency-dependent phase offsets that had not been fully removed during self-calibration, giving the appearance of a slight spatial shift of the continuum emission. This may also contribute to the very small velocity gradient shift in the \coone\ absorption-line centroid shown in Figure~\ref{fig:colineabsfit}.} in Figure~\ref{fig:emission}. The bulk of the Xa1 line profile excess would be inconsistent with the essentially unresolved excesses in the X5 data set.

In addition to the above, detection of a CND traced by \coone\ on such scales is improbable based on the spectral behavior. Positive features in the X5 and Xa1 line profiles are not sufficiently consistent or symmetrical, and nor are they well centered about \vsys. Some of that problem may arise from limited frequency coverage for $\nu>115.5$ GHz ($\vlos-\vsys < -1900$ \kms) and plausible overcorrection of the deep O$_2$ absorption. Finally, the maximum excess flux densities in the integrated line profiles reach only $\sim$0.15\% of the continuum levels, matching the $\sim$0.1\% amplitude variation seen when testing bandpass stability over larger frequency ranges \citep{kameno14}.

\subsection{Bandpass Calibration Accuracy}
\label{app:bandpass}

The maximum SDR~$\sim 2000$ measured in individual channels for the Band 3 data sets corresponds to $\mathrm{min}(\tau_0) \sim 0.0002$, enabling the narrow CO absorption-line detection discussed in Section~\ref{sec:narrowcoprops}. The larger frequency response is set by the bandpass calibration, so broad line signatures are impacted by the bandpass calibration accuracy. \citet{kameno14} demonstrated that cross-correlation bandpass calibration remained stable at the 0.2\% level over short time frames ($\sim$1 hr for Bands 3 and 6), producing a mostly flat spectral response  with a lower effective SDR~$\sim 500$ or $\min(\tau_0) \sim 0.002$ over an entire spw. Below this threshold, noticeable spectral structure arises above the white-noise predictions for frequency separations $\Delta\nu \gtrsim 250$ MHz (100$-$200 MHz and $\gtrsim$500 MHz) in Band 3 \citep[Band 6;][]{kameno14}. This means that fully calibrated spectra may have very low-amplitude fluctuations imprinted on velocity scales of $\Delta v \gtrsim 650$ \kms\ (130$-$160 \kms\ and $\gtrsim$650 \kms).

\begin{figure}[!ht]
    \centering
    \includegraphics[width=\textwidth]{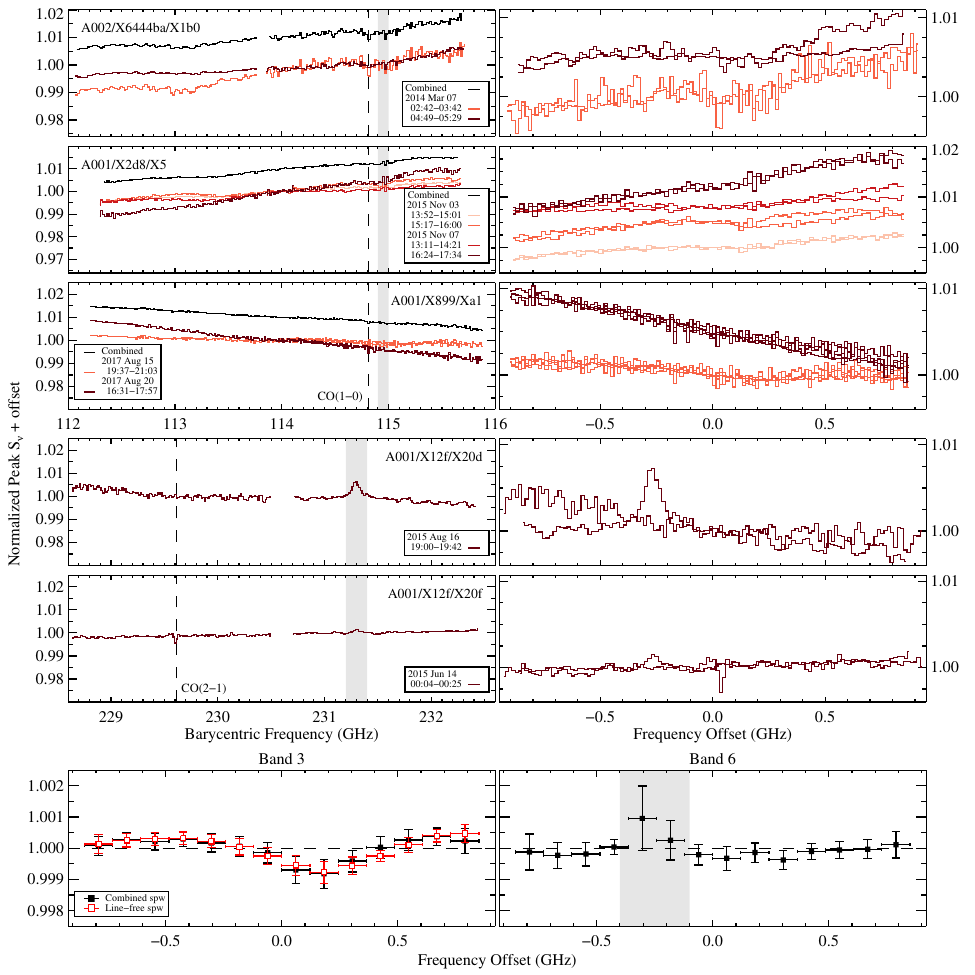}
    \caption{Bands 3 and 6 peak spectra of the M87 nucleus covering the redshifted \coone\ and \cotwo\ transitions in the upper sideband (\textit{top} and \textit{middle left panels}). Channels were averaged during \texttt{tclean} imaging to better determine broad spectral behavior, which appears to undulate well beyond the relative $-1000 < \vlos - \vsys < 100$ \kms\ range where the broad spectral deficit appears (see Figure~\ref{fig:broadabs}). Data were imaged within each UT time range and normalized at either 114 or 230 GHz. The combined peak spectra for each member OUS are offset slightly higher for clarity. Residual O$_3$ contamination is indicated (shaded regions). Comparing the spectral behavior for all spws in that sideband (\textit{top} and \textit{middle right}; grouped by UT date) reveals an apparent depression between 0--0.5 GHz at the 0.2--0.3\% level, even when a spw does not cover \coone. After dividing all these data by the corresponding full-spw or sideband continuum fit, the average Band 3 spectral response in the upper sideband shows clear correlated depressions at the 0.1\% level beyond the $1\sigma$ scatter of normalized intensities (\textit{bottom left}). In Band 6, the normalized response appears more flat (\textit{bottom right}).}
    \label{fig:bandpass}
\end{figure}

In Figure~\ref{fig:bandpass}, we compare the spectral behavior for individual spws in a single sideband from all standard ALMA programs covering redshifted \coone\ and \cotwo\ transitions. For Band 3 peak spectra, both the CO-centered and adjacent spws show similar fluctuations reaching the 0.1$-$0.3\% level relative to the spw channel outskirts. In the lower sideband, adjacent spws also show similar spectral response (e.g., for the X5 and X2be data; see Figure~\ref{fig:contabs}.) The downward fluctuation between 0$-$0.5 GHz relative to the spw center coincides with the apparent spectral deficits seen in Figure~\ref{fig:broadabs}. Such fluctuations arising from bandpass amplitude calibration may mimic faint, broad absorption for a fast outflow. For very low-amplitude, broad spectral features, we find a more appropriate SDR~$\sim 300-1000$ limit across an entire Band 3 spw using the standard bandpass calibration. Because these Band 3 spectra abut a strong O$_2$ absorption feature, the higher-frequency spw may also be negatively impacted by the $T_\mathrm{sys}$ undercorrection in the line wings as well as slight changes in $N_\mathrm{O_2}$ between the bandpass and science scans, either due to temporal variations or angular separation. In Band 6, the spectral shape appears more consistent with fluctuations at or below the 0.05\% level, matching its native SDR. However, the small sample size of deep Band 6 imaging with fine FDM binning prevents any clear conclusions.

We find further support for a bandpass calibration origin for these low-amplitude spectral features in the peak $S_\nu$ spectra in roughly contemporaneous observations shown in Figure~\ref{fig:contabs}. For example, the $S_\nu$ behavior over 102$-$104 GHz in the X5 data set is inconsistent with those covering the same spectral range in Xbe and X135 data sets, which were obtained just weeks apart. While the latter two member OUS used a different bandpass calibrator (J1229+0203) than the former (J1256--0547), differences in the calibrated, intra-spw shape is still within the quoted limitations for bandpass calibration \citep{privon24}. For these reasons, we view the very low amplitude spectral deficits or excess features in Figures~\ref{fig:broadabs} and \ref{fig:emission} as spurious and arising from a patterned response in the bandpass amplitude calibration for adjacent basebands. The mostly regular shape and extent of the seemingly broad \coone\ absorption in Figure~\ref{fig:broadabs} would then be a product of spw placement relative to the redshifted CO transition and proximity to the O$_2$ line that defines the end of the Band coverage.

\bibliography{main}{}
\bibliographystyle{aasjournal}

\end{document}